\documentclass[twocolumn,english,prl,showpacs,preprintnumbers,superscriptaddress]{revtex4-2}

\usepackage[utf8]{inputenc}
\usepackage{color}
\usepackage{amsmath,amssymb}
\usepackage{amssymb}
\usepackage{graphicx,pstricks}
\usepackage{textcomp}
\usepackage{dcolumn}
\usepackage{bm}
\usepackage[right]{eurosym}
\usepackage{float}
\usepackage[english]{babel}
\usepackage{blindtext}
\usepackage{babel}
\usepackage[normalem]{ulem}
\usepackage{wasysym}
\usepackage{hyperref}

\mathchardef\mhyphen="2D

\begin{document}
\title{Tracking Microhydration of the NaCl Rocksalt Molecule in Helium Nanodroplets by Penning Ionization Electron Spectroscopy}

\author{Ltaief Ben Ltaief}
\affiliation{Department of Physics and Astronomy, Aarhus University, 8000 Aarhus C, Denmark}

\author{Keshav Sishodia}
\affiliation{Quantum Center of Excellence for Diamond and Emergent Materials and Department of Physics, Indian Institute of Technology Madras, Chennai 600036, India}

\author{Robert Richter}
\affiliation{Elettra-Sincrotrone Trieste, 34149 Basovizza, Trieste, Italy}

\author{Mart\'i Pi}
\affiliation{Departament FQA, Facultat de F\'isica, Universitat de Barcelona, Av. Diagonal 645, 08028 Barcelona, Spain}
\affiliation{Institute of Nanoscience and Nanotechnology (IN2UB), Universitat de Barcelona, Barcelona, Spain}

\author{Manuel Barranco}
\affiliation{Departament FQA, Facultat de F\'isica, Universitat de Barcelona, Av. Diagonal 645, 08028 Barcelona, Spain}
\affiliation{Institute of Nanoscience and Nanotechnology (IN2UB), Universitat de Barcelona, Barcelona, Spain}

\author{Jussi Eloranta}
\affiliation{Department of Chemistry and Biochemistry, California State University at Northridge, Northridge, CA 91330, USA}

\author{Sivarama Krishnan}
\affiliation{Quantum Center of Excellence for Diamond and Emergent Materials and Department of Physics, Indian Institute of Technology Madras, Chennai 600036, India}

\author{Florent Calvo}
\affiliation{Universit\'e Grenoble Alpes, CNRS, LIPhy, 38000 Grenoble, France}

\author{Marcel Mudrich}
\affiliation{Department of Physics and Astronomy, Aarhus University, 8000 Aarhus C, Denmark}
\affiliation{Institute of Physics, University of Kassel, 34132 Kassel, Germany}


\date{October 2025}

\begin{abstract}
The microhydration of rock salt (NaCl) molecules was investigated using high-resolution Penning ionization electron spectroscopy (PIES) in helium nanodroplets. Although model calculations predict that NaCl molecules are fully submerged inside the droplets, PIES of NaCl are highly resolved, in stark contrast to other molecular species. Co-doping the droplets with a controlled number of $n=5$--10 water molecules
leads to efficient quenching of the NaCl Penning ionization signal and to its full suppression for $n\gtrsim 30$.
Accompanying density-functional theory (DFT) and force field calculations reveal a transition from contact ion pair structures to solvent-separated ion pairs at $n=12$--15. However, it takes $n\approx 17$ water molecules to form a complete solvation shell around the Cl$^-$ anion and as many as $n\approx 34$ to fully hydrate the Na$^+$ cation, thus the entire NaCl molecule, which rationalizes the experimental findings.
\end{abstract}

\maketitle

\section{Introduction}

The solvation of common rocksalt (NaCl) molecules in water
forming separated ion pairs is one of the most important processes in
chemistry that is of great relevance for biology, marine and
atmospheric chemistry, and everyday life. Although this process has
been extensively investigated both experimentally and theoretically,
open questions still remain regarding the microscopic structures and
energetics of the hydration complexes. Perhaps the most pertinent
question concerns the minimum number of water molecules necessary to
induce dissociation and charge separation of the Na$^+$Cl$^-$ ion
pair. A solvent-separated ion pair (SSIP) is characterized by solvent
molecules arranging themselves between the ions such that Na$^+$ and
Cl$^-$ each have their own hydration shell. As a result, the
interionic distance increases drastically compared with that of the
undissociated contact ion pair (CIP).

A large body of theoretical studies on this problem is available in
the literature, see~\cite{hou2017emergence} and references
therein. In particular, and as a response to the question ``How many water
molecules are necessary to dissolve a rock salt molecule?'', Jungwirth concluded that (H$_2$O)$_6$ represents the smallest water cluster capable of dissolving NaCl, based on earlier studies from the 1990s~\cite{jungwirth2000many}. However, the transition from CIP to SSIP ion pairs is all but sharp with increasing number $n$ of water molecules. According to the generally accepted picture, CIP structures dominate for $n < 4$; for $n\approx4$--6, one or more water molecules insert between the Na$^+$ and Cl$^-$ ions but still coordinate both ions simultaneously, thus forming so-called solvent-shared ion pairs; for $n\geq 6$--9, SSIP structures where each ion has its own hydration shell become more and more abundant.
Although the cluster calculations provide accurate energies and structures, the adopted models are not sufficiently large to represent the hydration structures in solution, especially for larger sizes where each ion becomes completely hydrated.

To address this question experimentally, NaCl(H$_2$O)$_n$ clusters containing a variable number of water molecules, $n=1,\,2,\dots$ have previously been
studied by infrared spectroscopy~\cite{tandy2016communication} and
photoelectron spectroscopy~\cite{hou2017emergence}. In the former experiment, Tandy~\textit{et al.} used superfluid helium (He) nanodroplets to isolate single NaCl molecules to which H$_2$O molecules were added one by one up to $n=4$. For $n=1$--3, the spectra
were consistent with the formation of the lowest-energy CIP structures in which each water molecule forms a single ionic hydrogen bond to an intact Na$^+$Cl$^-$ CIP; for $n = 4$, indications for the coexistence of at least two isomers were found. In a later experiment, Hou~\textit{et al.} studied NaCl$^-$(H$_2$O)$_n$ anions~\cite{hou2017emergence}, for which the SSIP-type structures appeared at $n = 2$. Neutral NaCl(H$_2$O)$_n$, which they studied only theoretically, occurred predominantly in CIP structures for $n < 9$; for $n=9$--12, CIP and SSIP
structures were found to be nearly degenerate in energy, largely in agreement with the generally accepted picture and matching the H$_2$O:NaCl molar ratio $\approx 9$ of NaCl-saturated solution~\cite{gregoire1998nai}.

In the current study, we applied a novel method --- helium nanodroplet Penning ionization electron spectroscopy --- to study the microsolvation of NaCl molecules by water molecules up to complete hydration. Helium nanodroplets (HNDs) are exceptional cryogenic matrices, offering an ultracold (0.37~K) and minimally perturbing environment ideal for the synthesis and spectroscopic investigation of individual molecules and molecular aggregates. The superfluid nature of HNDs facilitates the capture and aggregation of atoms and molecules as dopants, enabling the study of fundamental chemical and dynamical interactions that are often obscured in bulk phases or warmer gas-phase experiments. Consequently, HNDs are being used to explore molecular processes ranging from reaction dynamics at ultralow
temperatures~\cite{MuellerPRL:2009,krasnokutski2016ultra,mani2019acid,albrechtsen2023observing}
to the formation and characterization of exotic molecular
complexes~\cite{nauta2000formation,gutberlet2009aggregation,alevskovic2023nanostructured}.

Recently, we have expanded the scope of spectroscopic methods applied to HNDs to extreme ultraviolet (XUV) photoionization spectroscopy combined with electron and ion detection~\cite{Buchta:2013,BuchtaJCP:2013,laforge2024interatomic}. This approach has revealed a variety of intriguing interatomic and intermolecular relaxation processes involving the transfer of charge and energy between neighboring units and even over nanometer
distances~\cite{LaForgePRL:2016,LaForge:2019,ltaief:2023,ltaiefPRR:2024,bastian2024observation,laforge2024interatomic}. In particular, interatomic Coulombic decay (ICD) of atomic or molecular
dopants attached to HNDs induced by resonant excitation of the droplet, a process traditionally termed Penning ionization~\cite{Ltaief:2019}, turns out to be highly efficient in producing dopant ions and electrons~\cite{Buchta:2013,Shcherbinin:2018,Ltaief:2020,Mandal2020,asmussen2023secondary,sen2024electron,laforge2024interatomic,ltaief2025interatomic}. A recent study showed that a similar ICD process initiated  by electronic excitation reveals detailed information about the electronic structure of the water solvation shell surrounding the solvated cation in aqueous solutions~\cite{dupuy:2024}. However, Penning ionization electron spectra (PIES) in HNDs tend to be broad and rather featureless for most dopant
species~\cite{Shcherbinin:2018,Mandal2020,sen2024electron,asmussen2023secondary}; only for alkali metal atoms and clusters -- weakly bound to the surface of HNDs -- could well-resolved
PIES~\cite{Buchta:2013,Ltaief:2019,ltaief2025interatomic} be obtained.

Here we measure PIES of HNDs doped simultaneously with NaCl and H$_2$O molecules to track the microhydration dynamics of NaCl. The surprisingly high resolution of the PIES allows us to clearly distinguish Penning ionization of NaCl from that of bare sodium, which is also present in small amounts. Our main finding, namely that the full hydration of a NaCl molecule requires $n\approx 30$ water molecules, is supported by density-functional theory (DFT) and force field (FF) calculations. While the former approach confirms that the NaCl bond becomes significantly destabilized for $n\geq 6$, the FF exploration shows that complete hydration of both Na$^+$ and Cl$^-$ ions requires additional water molecules; caging of the solvent-dissociated Na$^+$ and Cl$^-$ ions remains incomplete up to $n=15$, even though the distance between the two ions is large enough to accommodate the hydrogen bond network in the SSIP fashion. Our simulations also reveal that the two Na$^+$ and Cl$^-$ ions exhibit different hydration patterns, reaching almost complete hydration shells for $n=34$ molecules for the cation, but only $n=17$ for the anion, while the Na$^+$--Cl$^-$ ionic bond is already well dissociated.

  
\section{Results and discussion}

\begin{figure}[t!]
	\center
        \includegraphics[width=0.9\columnwidth]{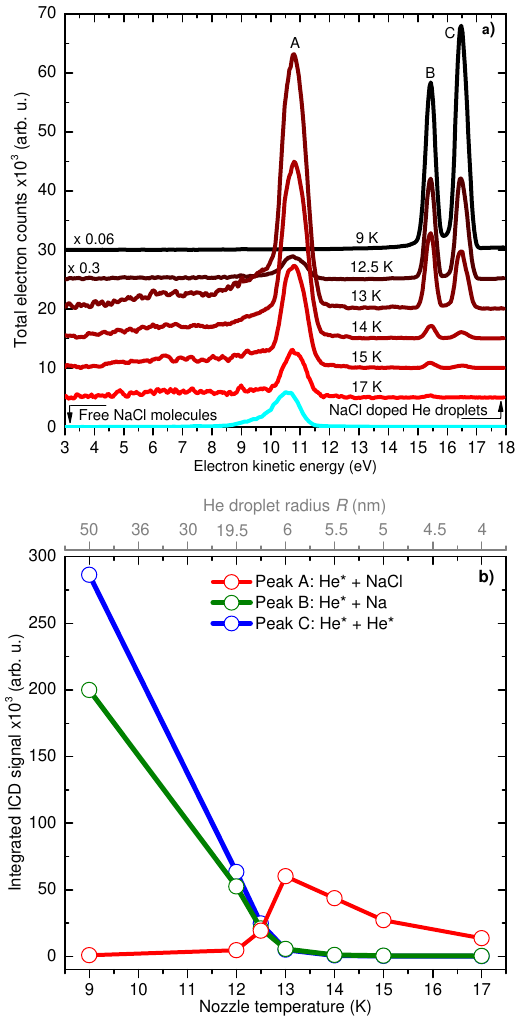}\caption{\label{fig:PIES-Tdep}
          a) Penning ionization electron spectra of NaCl-doped He nanodroplets recorded at a photon energy $h\nu = 21.6$~eV. The temperature of the NaCl doping cell was $460^\circ$C and the temperature of the He nozzle is given in the figure. The cyan line shows a reference spectrum measured for an effusive beam of NaCl molecules. The peak labeled A is due to Penning ionization of NaCl molecules. The peaks B and C are due to Penning ionization of Na and He$^*$ atoms, respectively;
          b) 
          peak integrals as a function of He nozzle temperature. The top horizontal scale indicates the estimated He droplet radius according to Ref.~\cite{Toennies:2004}.
          }
\end{figure}
We began our experimental study by recording electron spectra of NaCl-doped HNDs in the absence of water to investigate the efficiency of Penning ionization of NaCl, the structure of the resulting electron spectra, and their dependence on the HND size.
Fig.~\ref{fig:PIES-Tdep} a) shows the spectra of all electrons emitted from NaCl-doped HNDs for various temperatures of the He nozzle between 9 and 17~K (red and black lines). The corresponding HND radius is estimated to range from $R = 4$ to $50$~nm, respectively, by comparing to literature data~\cite{Toennies:2004}. The photon energy was set to $h\nu = 21.6$~eV to resonantly excite HNDs to their 2p state, thereby inducing Penning ionization of dopant atoms and molecules. All these spectra were measured at a temperature of the NaCl doping cell of $T_\mathrm{NaCl}= 460^\circ$C at which the NaCl Penning ionization signal was the highest, see Fig.~S1 and and Fig.~S2 b) in the SM. The peaks exhibit three distinct features denoted as A, B, and C, whose heights vary relative to one another as the HND size $R$ is changed. The dependence of peaks A, B, C on the temperature of the NaCl doping cell and thus on the number of NaCl embedded in the HNDs is presented in the Supplementary Material (SM), Fig.~S1 and Fig.~S2. 

Peak A is due to Penning ionization of NaCl molecules through the reaction
\begin{equation}
\label{eq:He-ICD}
\mathrm{He}^* + \mathrm{NaCl} \rightarrow \mathrm{He} + \mathrm{[NaCl]}^+ + e^-
\end{equation}
taking place inside HNDs. The characteristic kinetic energy of the electrons $e^-$ generated by this process, $E_e\approx 10.8$~eV, is approximately given by the difference between the energy stored ($E_\mathrm{He^*} =20.6$~eV) in the excited HNDs, which relax into the 2s\,$^1$S state prior to Penning ionization~\cite{Ltaief:2019,Mudrich:2020,ltaiefPRR:2024}, and the ionization potential of NaCl molecules (9.8~eV~\cite{potts1977photoelectron}). The narrow peak B results from Penning ionization of Na atoms attached to the surface of HNDs~\cite{BuchtaJCP:2013, Ltaief:2019, Asmussen:2021}. The peak position exactly matches the expected value, $E_e=E_\mathrm{He^*}-E_i^\mathrm{Na}=15.5$~eV, where $E_i^\mathrm{Na}=5.1$~eV is the atomic ionization energy of Na. These Na atoms are present at low concentration due to contamination of the vacuum chamber from previous experiments. Here, peak B serves as a reference for PIES of dopants attached to the HND surface~\cite{Buchta:2013,Ltaief:2019,ltaief2025interatomic}. Peak C is due to a Penning ionization process that involves two He$^*$ atoms in the 2s\,$^1$S state excited in large HNDs by two-photon absorption~\cite{ltaiefPRR:2024}. Accordingly, its position is at $E_e=2\times E_\mathrm{He^*}-E_i^\mathrm{He}=16.6$~eV, where $E_i^\mathrm{He}=24.6$~eV is the ionization energy of He atoms. This feature is a sensitive probe for the size of the HNDs~\cite{ltaiefPRR:2024}. The integrals of peaks A, B, and C are shown in Fig.~\ref{fig:PIES-Tdep} b) as red, green, and blue symbols, respectively.

Peak A starts to appear as the He nozzle is cooled below $T_\mathrm{nozzle}= 20$~K where condensation of He into nanodroplets sets in. It reaches a maximum at $T_\mathrm{nozzle}=13$~K (HND radius $R\approx6$~nm) before sharply dropping and vanishing as the nozzle is cooled further down to $T_\mathrm{nozzle}\leq9~$K ($R\gtrsim50$~nm). The signal rise in the range $T_\mathrm{nozzle}=17$--13~K is due to the increase of both the flux and the size of HNDs, where the latter leads to an increase of the probability of the droplets to pick up one or more NaCl molecules. The sharp drop at $T_\mathrm{nozzle}\leq9~$K, which is observed for all dopant species that are submerged in the interior of the HNDs~\cite{asmussen2023dopant}, reflects the finite effective range of Penning ionization as the He$^*$ excitation tends to form a void bubble that emerges to the droplet surface. There, the He$^*$ remains weakly bound to the HND in a shallow dimple, spatially separated from the submerged dopant~\cite{sishodia2025xuvfluorescenceprobeinteratomic}. Accordingly, peaks B and C keep rising as the HNDs grow larger at $T_\mathrm{nozzle}\leq13~$K; surface-bound Na atoms and He$^*$ atoms likely roam about the HND surface and eventually decay by Penning ionization once they collide~\cite{sishodia2025xuvfluorescenceprobeinteratomic}. The emitted electron can reach the detector without scattering with He atoms surrounding the dopant. In contrast, electrons emitted from submerged dopants undergo massive scattering, leading to a broadening of the peaks in the electron spectra~\cite{asmussen2023electron}; large HNDs with $R\gtrsim 20$~nm even tend to trap electrons thereby suppressing their detection~\cite{asmussen2023dopant}. This likely explains the complete vanishing of peak A at 9~K nozzle temperature. The strongly differing HND size dependences of the NaCl PIES vs. Na PIES clearly indicate that NaCl is submerged inside HNDs, as expected for most molecular species.

For reference, the cyan line in Fig.~\ref{fig:PIES-Tdep}a) shows the photoelectron spectrum of free NaCl molecules prepared in an effusive beam and photoionized at $h\nu = 20.6$~eV in a separate experiment. It contains only one peak corresponding to ionization from the Cl 3p-derived valence band. The fact that peak A in the PIES has nearly the same position and width as the photoline of effusive NaCl highlights the excellent spectral resolution achieved with PIES for this particular system. In this respect, NaCl resembles more the surface-bound Na and He$^*$ atoms, see peaks B and C and Refs.~\cite{Ltaief:2019,ltaiefPRR:2024,ltaief2025interatomic}, as opposed to molecules submerged inside HNDs which mostly feature broad, nearly structureless PIES~\cite{Shcherbinin:2018,Mandal2020,sen2024electron,asmussen2023secondary}.

To resolve the question of the location of the NaCl molecule inside HNDs or at the surface, we carried out dedicated model calculations. We used two complementary methods based on either static $^4$He density-functional theory (DFT) continuum-like calculations or atomistic path-integral molecular dynamics (PIMD). Both methods, which are described in the SM, predict that the NaCl molecule should be located near the center of the HNDs (see Fig.~S4 in the SM). The $^4$He-DFT calculations indicate significant localization of He atoms in the vicinity of the Na$^+$ cation and the Cl$^-$ anion, suggestive of a snowball being formed surrounded by the remainder of the HND acting as a quantum solvent. The PIMD simulations confirm this picture as well as the asymmetric character of the snowball with enhanced He localization near the cation relative to the anion. It is possible that NaCl dopants occupy various positions inside larger HNDs with a nonzero probability of localizing near the HND surface, similar to alkaline-earth-metal atoms~\cite{ren2007surface} and indium atoms~\cite{thaler2020ultrafast}, likely with the more heliophilic Na$^+$ ion end pointing toward the droplet center and the Cl$^-$ ion pointing toward the surface.


Having established PIES as a sensitive and precise method for probing NaCl molecules despite their location inside HNDs, we now turn to the investigation of their microsolvation by water. To this end, we add a controlled number of H$_2$O molecules to the NaCl-doped HNDs. The presence of H$_2$O molecules introduces hydrogen-bonding interactions, facilitating the formation of hydrated NaCl complexes within the HND environment. As the polar water molecules are expected to interact much more strongly with the Na$^+$ and Cl$^-$ ions than with He atoms, it is anticipated that they will efficiently aggregate around the NaCl molecule, thereby significantly altering its susceptibility to Penning ionization.

\begin{figure}[t!]
	\center
        \includegraphics[width=0.95\columnwidth]{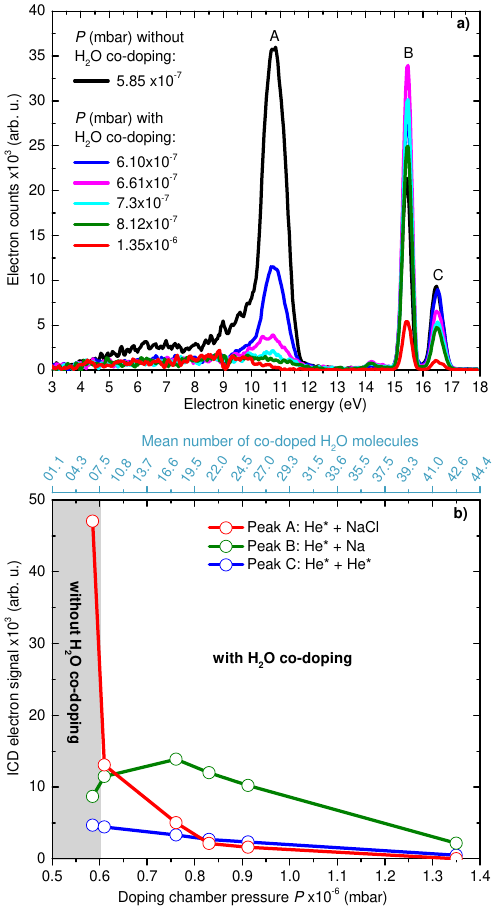}\caption{\label{fig:PIES-H2O}a)
          PIES of NaCl-doped He nanodroplets co-doped with H$_2$O molecules at variable vapor pressure in the doping chamber. The He nozzle temperature was set to 13~K, the NaCl oven temperature was $T_\mathrm{NaCl}=460^\circ$C, and the photon energy was $h\nu = 21.6$~eV. The spectrum in black was recorded in the absence of H$_2$O co-doping; b) peak integrals as a function of water co-doping pressure. The axis at the top of panel b) denotes the mean number of co-doped H$_2$O molecules per He nanodroplet, as estimated using the formulas described in SM section 2.}
\end{figure}

Fig.~\ref{fig:PIES-H2O}a) shows PIES measured for optimal NaCl doping conditions ($T_\mathrm{NaCl}=
460^\circ$C) and for various levels of co-doping the HNDs with H$_2$O molecules. The top horizontal axis shows the estimated number of co-doped H$_2$O molecules per HND. Most strikingly, by leaking small amounts of water vapor in the doping cell such that the doping chamber pressure rises from 5.8 to $6.1\times 10^{-7}$~mbar, the NaCl Penning ionization signal steeply drops to less than half of its previous intensity, whereas the Na and He$^*$ signals (peaks B and C) remain almost constant. The estimated number of co-doped H$_2$O molecules responsible for this drastic drop of the Penning ionization of NaCl is about 7 based on the model described in Refs.~\cite{kuma2007laser, De:2024}. As the H$_2$O pressure is further increased, the NaCl signal in the PIES further drops and vanishes almost completely when the number of H$_2$O co-dopants reaches around 30, see Fig.~\ref{fig:PIES-H2O}b). In stark contrast, in the same range of H$_2$O co-doping the intensity of peak B (Na Penning ionization) slightly increases to reach a maximum at about 17 H$_2$O molecules before gradually decreasing at higher H$_2$O co-doping levels, while peak C (He$^*$ Penning ionization) only slowly decreases. The opposing trends of dropping NaCl signal and rising Na signal in the range of low co-doping (0.6--$0.7\times 10^{-6}$~mbar) reflect the competition between Penning ionization of NaCl and Na dopants; as Penning ionzation of NaCl becomes less efficient due to microhydration, more He$^*$ are available in the HNDs to induce Penning ionization of Na. The overall drop of all signals occurs only at higher doping levels, due to shrinking of the HNDs caused by additional scattering with H$_2$O molecules in the doping chamber, which eventually leads to the destruction of the HND beam. This clearly indicates that the suppression of NaCl Penning ionization is mainly caused by the binding of the water molecules to the NaCl molecules, rather than by HND beam destruction.

\begin{figure}[t!]
	\center
        \includegraphics[width=1.\columnwidth]{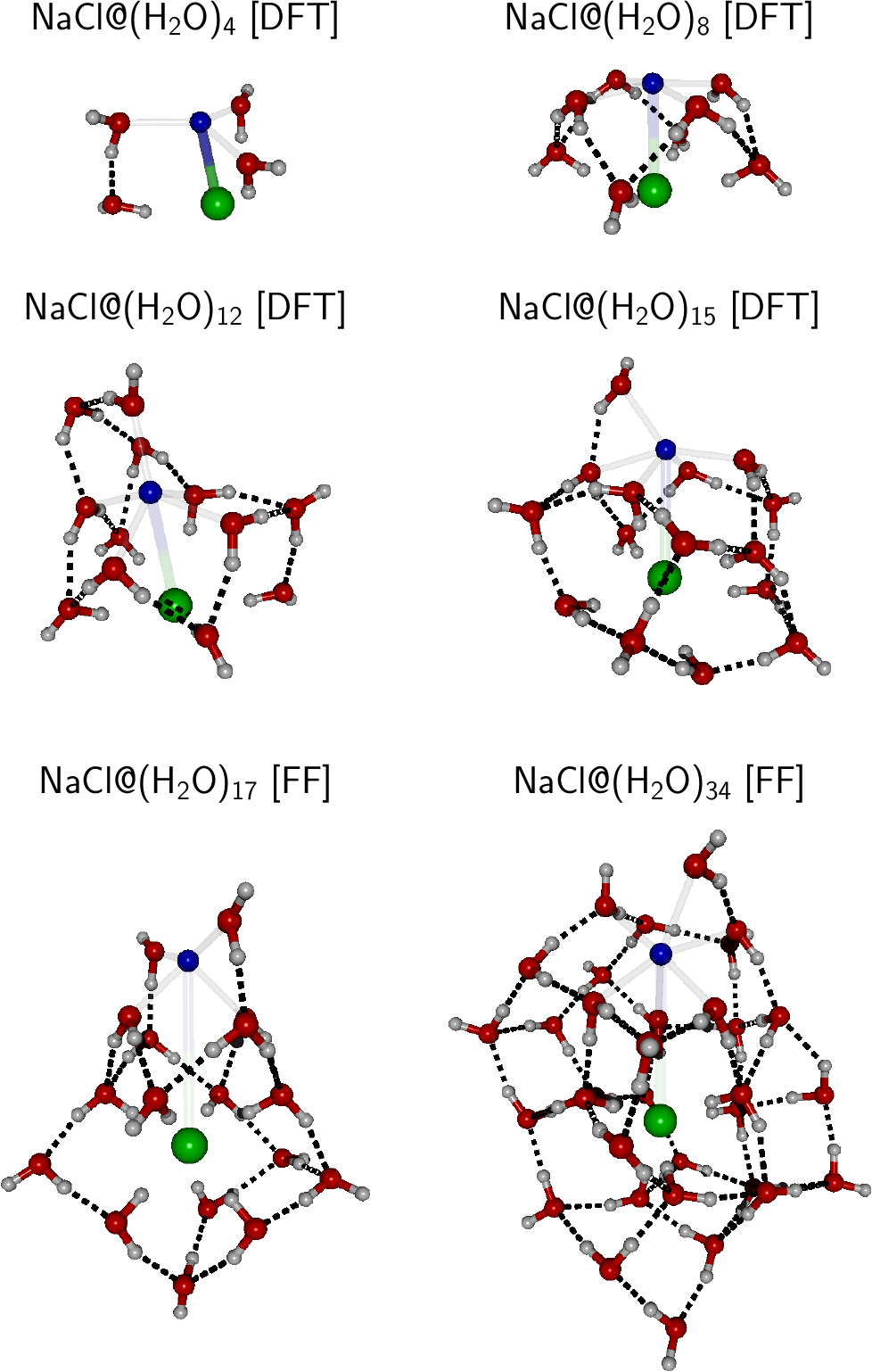}\caption{\label{fig:structures}
          Lowest-energy structures of selected clusters obtained from DFT calculations at small sizes, or from the force field (FF) exploration at larger sizes.}
\end{figure}

To rationalize these experimental findings, candidates for the stable structures of NaCl(H$_2$O)$_n$ clusters were explored using a computational methodology combining force field exploration and DFT reoptimization, as detailed in the Theoretical Modeling section. To better explore the competing CIP and SSIP regions of the energy landscape, the distance between the Na$^+$ and Cl$^-$ ions was restricted to dedicated values through a dedicated (umbrella) harmonic potential, ranging from 2~\AA\ to 6~\AA\ by steps of 0.5~\AA. In clusters containing $n>15$ water molecules, only candidate structures were produced from the FF approach. Selected examples of the resulting putative global minima are shown in Fig.~\ref{fig:structures} for $n=4$, 8, 12, and 15. At small sizes, the hydration pattern identified in earlier studies~\cite{jungwirth2000many,hou2017emergence,shi23} is recovered, with the binding of the first water molecules to the Na$^+$ cation, and with the fourth added water molecule being hydrogen-bonded to an already existing water. 
However, even for $n=15$, both Na$^+$ and Cl$^-$ ions are still not fully hydrated, though we note that the water coordination of Cl$^-$ is greater than that of Na$^+$ at this size. Instead, these molecules create a hydrogen-bond network in which the Na$^+$ ion, then the Cl$^-$ ion, are successively caged.

\begin{figure}[t!]
	\center
        \includegraphics[width=1.\columnwidth]{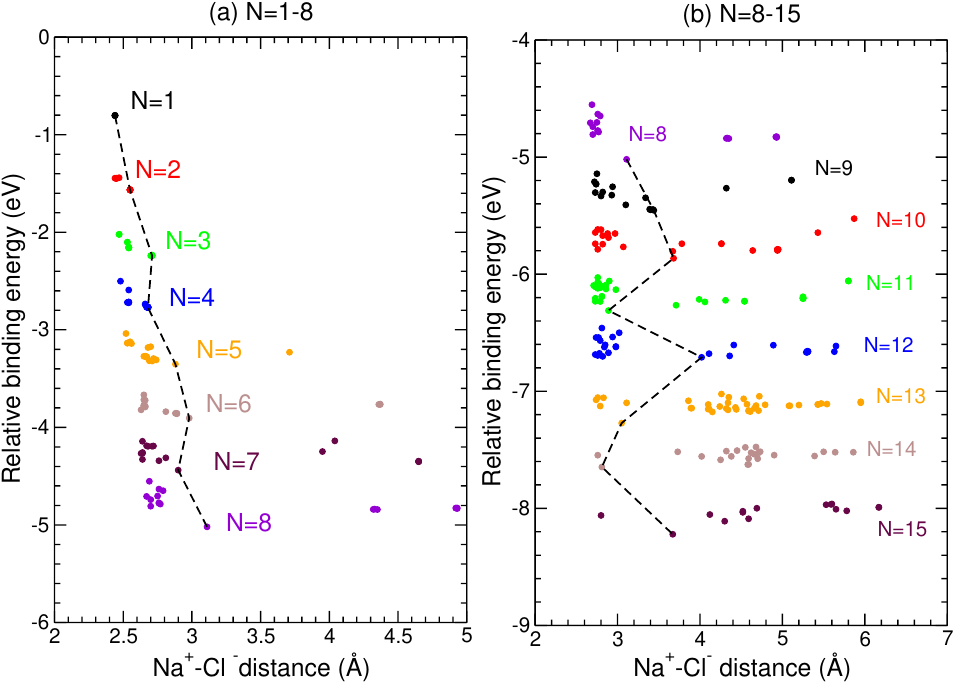}\caption{\label{fig:rbe}Relative binding energy of various configurations of the           NaCl(H$_2$O)$_n$ clusters obtained at the DFT level of theory, sorted according to the Na-Cl equilibrium distance. Each cluster size $n$ is associated with a different color, and the dashed lines join the lowest energy configurations found for all sizes.}        
\end{figure}

The destabilization effect of water molecules on the Na-Cl ionic bond was assessed in our calculation by determining a relative binding
energy (RBE) as
\begin{equation}
  \mbox{RBE}=E[\mbox{NaCl}({\rm H}_2{\rm O})_n] - nE[{\rm H}_2{\rm O}]
  -E[{\rm NaCl}],
  \label{eq:rbe}
\end{equation}
where $E[{\rm X}]$ denotes the total electronic energy of molecule X,
corrected for the zero-point vibrational energy contribution taken here simply in the harmonic approximation, and to correlate this quantity to the interatomic distance $R_{\rm NaCl}$, for the various structures minimized at the DFT level. The resulting values are shown in Fig.~\ref{fig:rbe}.

The trend found from these calculations matches the existing interpretation of a steady destabilization of the Na-Cl ionic bond as more and more water molecules are being added. The picture obtained at small sizes, in particular, is in good agreement with the recent work by Shi and coworkers~\cite{shi23}. In the range of larger sizes, $n=8$--15, significant fluctuations are noted with the strictly lowest-energy minima found in our samples to be of the CIP (non-dissociated) type at sizes 11, 13, and 14, whereas the lowest minima at the neighboring sizes of 12 and 15 belong to the SSIP category. Note, however, that our exploration of the energy landscapes for these larger clusters is probably incomplete, despite allowing for large variations of the values of $R_{\rm NaCl}$. 

\begin{figure}[t!]
	\center
        \includegraphics[width=1.\columnwidth]{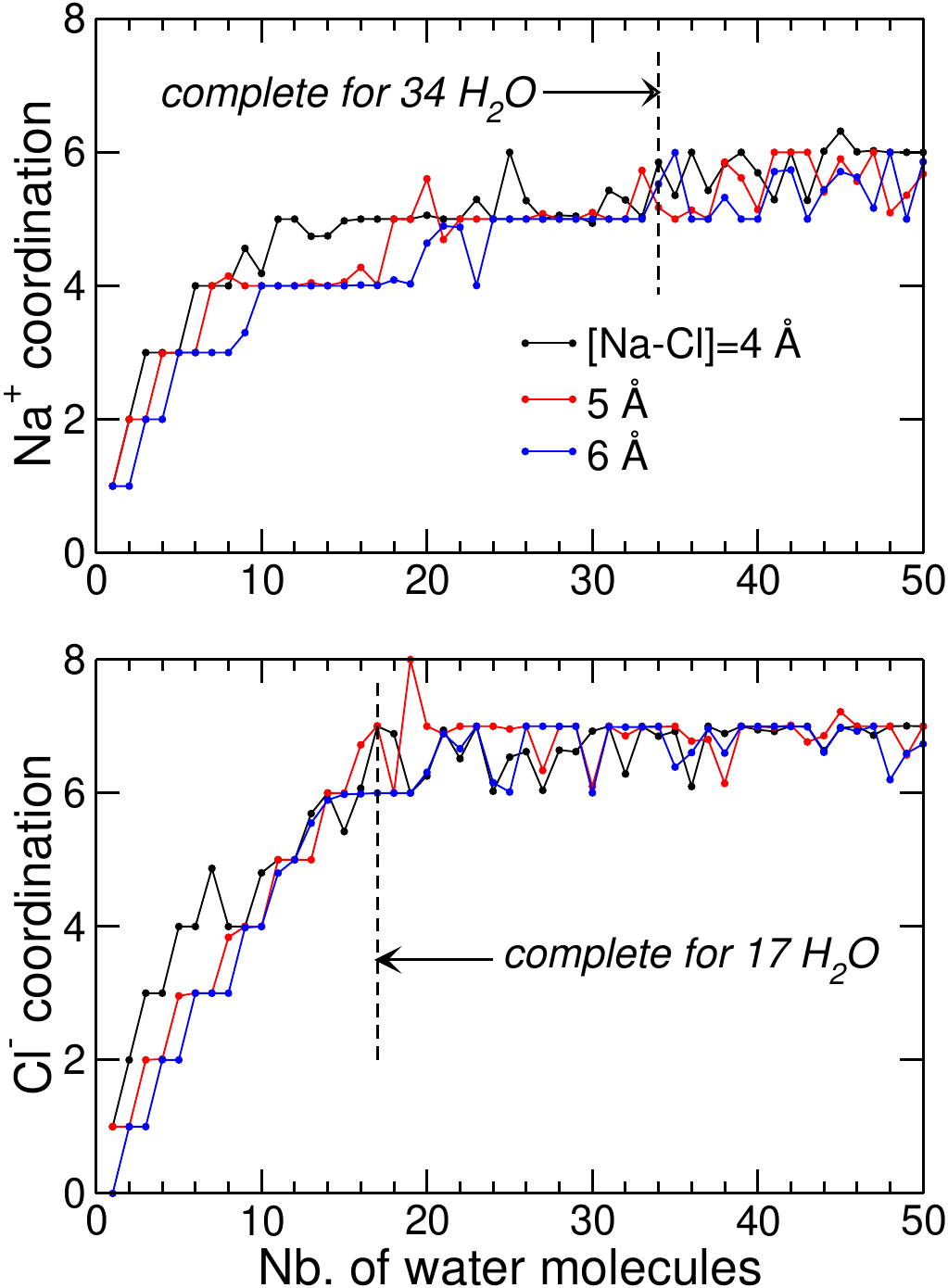}\caption{\label{fig:coord}
          Effective coordination numbers of the Na$^+$ (upper panel) and Cl$^-$ (lower panel) ions by water molecules in low-temperature samples of NaCl(H$_2$O)$_n$ clusters with
          the Amber {\em ff}99 force field, keeping the distance between the two ions close to 4, 5, or 6~\AA.}
\end{figure}

Finally, the coordination numbers specific to both ions were determined up to $n=50$, restricting the energy landscape exploration
at the force field level but imposing three values $R_{\rm NaCl}=4$, 5, or 6~\AA\ through the umbrella potential. The resulting coordination numbers are represented in Fig.~\ref{fig:coord} as an increasing function of $n$.
From these results, the Cl$^{-}$ anion reaches its maximum coordination number of 7 once 17 water molecules are present. However, Na$^+$ only accommodates at most 6 water molecules, and does so consistently for at least $n=34$ and above. Thus, our theoretical exploration of structures, which exceeds earlier quantum chemical investigations, reveals that the hydration pattern does not proceed in a fully regular manner with increasing numbers $n$ of water molecules; only at small sizes $n<4$ do the first solvent molecules bind to the  Na$^{+}$ ion. Between $n=5$ and 8, the additional molecules bind to the Cl$^{-}$ ion and to the existing waters through hydrogen bonds. Between 9 and approximately 15 molecules, the additional water molecules successively bind partly to the cation, but more importantly to the anion. At $n=17$, Cl$^-$ is almost fully
hydrated, while Na$^+$ still misses one solvent molecule. Above $n=17$, the hydrogen bond network around Cl$^-$ grows further, without affecting much the local environment of the cation, until finally it becomes fully hydrated itself once $n=34$ is approximately reached. 

\section{Conclusions}

We have successfully employed Penning ionization electron spectroscopy in HNDs as a highly sensitive probe to investigate the microhydration of NaCl molecules under rather well-controlled conditions of complexation with water molecules. The NaCl electron spectra are surprisingly well-resolved, given that the NaCl molecule is fully submerged in the HND interior as predicted independently by continuum-based DFT and atomistic-based PIMD simulations. When a small number of 5--10 water molecules were added to the NaCl-doped helium nanodroplets, the Penning ionization electron signal of NaCl drastically dropped. This indicates that NaCl and water molecules efficiently interact inside the HND; apparently, the NaCl molecules are efficiently screened from the attacking He$^*$ in the HND and their structure is altered in such a way that their Penning ionization is quenched.


Accompanying model calculations combining force field exploration and structural refinement at the DFT level provide detailed insights into the structure of NaCl hydration complexes; in this way they give a more nuanced answer to the long-standing question of how many water molecules are required to dissolve a NaCl molecule. While the Na-Cl ionic bond is indeed significantly destabilized by as few as 4--6 water molecules, this hardly corresponds to complete hydration from the perspective of individual ions. Our theoretical results instead suggest a more gradual picture of the solvation process of NaCl into separated ion pairs: the Cl$^-$ anion reaches its maximum coordination number of approximately 7 at around 17 water molecules, whereas the Na$^+$ cation requires nearly twice as many — approximately 34 — waters to become fully hydrated. Thus, already the partial encapsulation of the NaCl molecule in the process of forming CIP structures effectively screens the NaCl core from interaction with a metastable He$^*$ atom generated by photoexcitation of the HND. However, full quenching of NaCl Penning ionization only occurs upon formation of SSIP complexes and full hydration of the Na$^+$ and Cl$^-$ ions. 

The positive correspondence between our structure calculations and the experimental findings that we have established with this work will allow for further studies on the microsolvation of other types of salts, metals, as well as their clusters and complexes. A particularly intriguing prospect is studying the microsolvation of atomic and molecular cations and possibly anions, which are amenable to HND Penning ionization if their ionization energy is sufficiently low, as it has recently beenshown~\cite{schobel2010sequential,Foitzik_Large_Ordered_Helium_Solvation_Shells}.
The synergy between the unique cryogenic spectroscopic environment of HNDs, which allows for the synthesis of specific hydrated complexes, and the sensitive probing of solvation effects provided by PIES, opens a new window onto unraveling fundamental solvation mechanisms at the molecular level. 

\section{Experimental setup}
The experiments were carried out using a He nanodroplet apparatus combined with a hemispherical electron analyzer (HEA, model VG-220i) that were mounted at the GasPhase beamline of the Elettra synchrotron facility in Trieste, Italy. This setup, which has previously been used only to study pure HNDs, is described elsewhere in detail~\cite{ltaief:2023,ltaiefPRR:2024}.
The advantages of the HEA are
(i) its high resolution of $<0.1$~eV; (ii) the possibility to focus onto a small range of the electron spectrum while suppressing electrons whose energies lie outside this range; and (iii) the selective detection of electrons emitted in a small acceptance volume. In this way, spurious electrons emitted from the residual gas along the photon beam and from stray light hitting spectrometer surfaces are suppressed. The magic-angle geometry is chosen to ensure that the measured yields of electrons are independent of their angular distributions. 

A continuous beam of HNDs containing between about $3\times10^3$ (droplet radius $R$ $\approx$ 4~nm) and $4\times10^7$ ($R$ $\approx$ 50~nm) He atoms per droplet was generated by expanding ultrahigh purity $^4$He gas out of a cryogenic nozzle at a temperature ranging from 17 down to 9~K, respectively, at a He backing pressure of 50~bar. The HNDs were doped first with NaCl molecules and then with H$_2$O molecules by passing the droplets through a radiatively heated vapor cell containing NaCl powder (10~mm length) and an adjacent gas cell (18~mm length), respectively. The temperature of the NaCl cell was varied in the range of $T_\mathrm{NaCl}= 420$--$515^\circ$C. At $T_\mathrm{NaCl}= 460^\circ$C, we estimate the mean number of NaCl molecules picked up by the He nanodroplet with radius $R\approx6~$nm to be $\approx1$~NaCl molecule, using the simple model described in Ref.~\cite{kuma2007laser}. The H$_2$O vapor pressure in the gas-doping cell was varied in the range 5.9--13.5$\times 10^{-7}$~mbar corresponding to an estimated number of water molecules picked up by the HNDs after doping with NaCl molecules of approximately 6 to 43, respectively. However, the exact number of NaCl and H$_2$O dopants picked up by the HNDs is difficult to estimate as the size of the HNDs shrinks significantly during the doping process~\cite{Buenermann:2011}.

A mechanical beam chopper was placed between the skimmer and the doping cell to discriminate the spectra of doped HNDs from the background due to photoionization of residual gas. In this study, the photon energy was set at $h\nu = 21.6$~eV to match the 1s$\rightarrow$2p absorption resonance of HNDs~\cite{Joppien:1993,Buchta:2013}. The use of a variable-angle spherical grating monochromator ensured narrow-band radiation with a photon flux $\Phi\approx 5\times 10^{11}~$s$^{-1}$. A tin (Sn) filter was inserted into the beamline to suppress higher-order radiation.

\section{Theoretical modeling}
\label{sec:th}

\subsection{Stable configuration of hydrated clusters}

The stable structures of NaCl(H$_2$O)$_n$ clusters have been investigated with quantum chemical methods in the past, and Shi~\textit{et al.}~\cite{shi23} have notably explored clusters containing up to $n=9$ water molecules. In order to extend this range, while keeping a reasonable computational cost, a two-stage conformational sampling procedure was implemented, allowing to produce realistic structures up to $n=15$. Here, it should also be kept in mind that, under the experimental conditions of co-doping the droplets with NaCl and, subsequently, the water molecules, and under the extreme cryogenic conditions provided by the He nanodroplet environment, it is likely that the structure of the hydrated clusters may deviate from the global minimum structure owing to the rapid freezing process. Such out-of-equilibrium processes are well known in the physical chemistry of molecular dopants in He droplets \cite{nauta1999}. However, here we are only concerned with neat NaCl(H$_2$O)$_n$ clusters in the gas phase, i.e. not explicitly embedded in HNDs.

In the first stage of our exploration, large numbers of relevant structures were produced directly at the molecular level, using a simple classical force field to describe the interactions between water molecules and the Na$^+$ and Cl$^-$ ions. Here we adopted Amber {\em
  ff}99  as the main workhorse for this stage, which notably relies on the simple flexible TIP3P model for interactions among water molecules \cite{wang00}. One explicit but important modification was made to this FF
by treating the interaction between Na$^+$ and Cl$^-$ ions with a
simple harmonic potential, mostly with the aim of covering broader
regions of the energy landscape by repeating optimizations for
different values of the Na-Cl distance $R_{\rm NaCl}$.

Replica-exchange molecular dynamics (REMD) simulations were performed for $n=1$--15 water molecules interacting with the Na$^+$ and Cl$^-$ ions at various values of their distance $R_{\rm NaCl}$. The computational details of the REMD simulations are provided as Supplementary Material.

In this second stage, up to 50 candidate structures produced using the
force field and after varying the Na-Cl distance were further locally
minimized using (electronic) density-functional theory, employing the M06-2X double hybrid functional together with the 6-311++G(d,p) functional in
all-electron calculations performed using the Gaussian16 suite of
programs \cite{g16}. This approach was found earlier\cite{mcbmuracil} to be accurate
and efficient for other microhydrated compounds. The structures
obtained from these minimizations were sorted based on their Na-Cl
distance (not constrained during the DFT optimization), and their
relative energetic stability, using the RBE quantity defined in Eq.~(\ref{eq:rbe}). With the current DFT method, we find $E[{\rm   NaCl}]=-622.528$ Hartree and $E[{\rm H}_2{\rm O}]=-76.399$~Hartree.

\subsection{Towards larger clusters and the complete hydration limit}

While it is not practical to undertake global optimization at the same
level of DFT quality for clusters containing more than 15 water
molecules, the FF approach with fixed Na-Cl distance was also employed to cover a much broader size range, $n=15$--50, but for the sole purpose of shedding light onto the possible hydrogen bonding network adopted under the constraint of a fixed interionic distance, thereby ensuring both CIP and SSIP regions of the energy landscape would be covered separately. The same computational methodology based on the REMD approach was thus followed, but instead of locally optimizing the resulting configurations, the average coordination of Na$^+$ and Cl$^-$ ions to the neighboring water molecules was determined from the instantaneous configurations sampled at the 100~K REMD trajectory. This temperature, which is markedly higher than the experimental conditions of HNDs, was chosen to increase the structural diversity and in the scrutinized sample. For an instantaneous configuration in the sample, coordination numbers for Na$^+$ and Cl$^-$ ions were calculated simply by counting the number of oxygen atoms within 3 and 3.2~\AA\ from the Na$^+$ and Cl$^-$ ions, respectively. A slightly larger cutoff for the anion was imposed to accommodate the fact that hydrogen atoms bind to it, rather than oxygen atoms.

\section{Acknowledgement}
L.B.L. acknowledges support by the Villum foundation via the Villum Experiment grant No. 58859. M.M. acknowledges support by the Novo Nordisk Foundation (grant no. NNF23OC0085401). S.R.K. thanks Dept. of Science and Technology, Govt. of India, for support through the DST-DAAD scheme and Science and Eng. Research Board. S.R.K. and K.S. acknowledge the support of the Scheme for Promotion of Academic Research Collaboration, Min. of Edu., Govt. of India, and the Institute of Excellence programme at IIT-Madras via the Quantum Center for Diamond and Emergent Materials. S.R.K. gratefully acknowledges support of the Max Planck Society's Partner group programme. S.R.K. gratefully acknowledges CEFIPRA (Indo-French Centre for the Promotion of Advanced Research) for generous support. This work was supported by Grant No. PID2023-147475NB-I00, funded by MICIU/AEI/10.13039/501100011033. The research leading to these results has been supported by the COST Action CA21101 ``Confined Molecular Systems: From a New Generation of Materials to the Stars (COSY)''.

\bibliography{NaCl_Bib}

\begin{thebibliography}{47}%
\makeatletter
\providecommand \@ifxundefined [1]{%
 \@ifx{#1\undefined}
}%
\providecommand \@ifnum [1]{%
 \ifnum #1\expandafter \@firstoftwo
 \else \expandafter \@secondoftwo
 \fi
}%
\providecommand \@ifx [1]{%
 \ifx #1\expandafter \@firstoftwo
 \else \expandafter \@secondoftwo
 \fi
}%
\providecommand \natexlab [1]{#1}%
\providecommand \enquote  [1]{``#1''}%
\providecommand \bibnamefont  [1]{#1}%
\providecommand \bibfnamefont [1]{#1}%
\providecommand \citenamefont [1]{#1}%
\providecommand \href@noop [0]{\@secondoftwo}%
\providecommand \href [0]{\begingroup \@sanitize@url \@href}%
\providecommand \@href[1]{\@@startlink{#1}\@@href}%
\providecommand \@@href[1]{\endgroup#1\@@endlink}%
\providecommand \@sanitize@url [0]{\catcode `\\12\catcode `\$12\catcode `\&12\catcode `\#12\catcode `\^12\catcode `\_12\catcode `\%12\relax}%
\providecommand \@@startlink[1]{}%
\providecommand \@@endlink[0]{}%
\providecommand \url  [0]{\begingroup\@sanitize@url \@url }%
\providecommand \@url [1]{\endgroup\@href {#1}{\urlprefix }}%
\providecommand \urlprefix  [0]{URL }%
\providecommand \Eprint [0]{\href }%
\providecommand \doibase [0]{https://doi.org/}%
\providecommand \selectlanguage [0]{\@gobble}%
\providecommand \bibinfo  [0]{\@secondoftwo}%
\providecommand \bibfield  [0]{\@secondoftwo}%
\providecommand \translation [1]{[#1]}%
\providecommand \BibitemOpen [0]{}%
\providecommand \bibitemStop [0]{}%
\providecommand \bibitemNoStop [0]{.\EOS\space}%
\providecommand \EOS [0]{\spacefactor3000\relax}%
\providecommand \BibitemShut  [1]{\csname bibitem#1\endcsname}%
\let\auto@bib@innerbib\@empty
\bibitem [{\citenamefont {Hou}\ \emph {et~al.}(2017)\citenamefont {Hou}, \citenamefont {Liu}, \citenamefont {Li}, \citenamefont {Xu}, \citenamefont {Gao},\ and\ \citenamefont {Zheng}}]{hou2017emergence}%
  \BibitemOpen
  \bibfield  {author} {\bibinfo {author} {\bibfnamefont {G.-L.}\ \bibnamefont {Hou}}, \bibinfo {author} {\bibfnamefont {C.-W.}\ \bibnamefont {Liu}}, \bibinfo {author} {\bibfnamefont {R.-Z.}\ \bibnamefont {Li}}, \bibinfo {author} {\bibfnamefont {H.-G.}\ \bibnamefont {Xu}}, \bibinfo {author} {\bibfnamefont {Y.~Q.}\ \bibnamefont {Gao}},\ and\ \bibinfo {author} {\bibfnamefont {W.-J.}\ \bibnamefont {Zheng}},\ }\bibfield  {title} {\bibinfo {title} {Emergence of solvent-separated \uppercase{N}a$^+$--\uppercase{C}l$^-$-ion pair in salt water: Photoelectron spectroscopy and theoretical calculations},\ }\href@noop {} {\bibfield  {journal} {\bibinfo  {journal} {J. Phys. Chem. Lett.}\ }\textbf {\bibinfo {volume} {8}},\ \bibinfo {pages} {13} (\bibinfo {year} {2017})}\BibitemShut {NoStop}%
\bibitem [{\citenamefont {Jungwirth}(2000)}]{jungwirth2000many}%
  \BibitemOpen
  \bibfield  {author} {\bibinfo {author} {\bibfnamefont {P.}~\bibnamefont {Jungwirth}},\ }\bibfield  {title} {\bibinfo {title} {How many waters are necessary to dissolve a rock salt molecule?},\ }\href@noop {} {\bibfield  {journal} {\bibinfo  {journal} {J. Phys. Chem. A}\ }\textbf {\bibinfo {volume} {104}},\ \bibinfo {pages} {145} (\bibinfo {year} {2000})}\BibitemShut {NoStop}%
\bibitem [{\citenamefont {Tandy}\ \emph {et~al.}(2016)\citenamefont {Tandy}, \citenamefont {Feng}, \citenamefont {Boatwright}, \citenamefont {Sarma}, \citenamefont {Sadoon}, \citenamefont {Shirley}, \citenamefont {Das Neves~Rodrigues}, \citenamefont {Cunningham}, \citenamefont {Yang},\ and\ \citenamefont {Ellis}}]{tandy2016communication}%
  \BibitemOpen
  \bibfield  {author} {\bibinfo {author} {\bibfnamefont {J.}~\bibnamefont {Tandy}}, \bibinfo {author} {\bibfnamefont {C.}~\bibnamefont {Feng}}, \bibinfo {author} {\bibfnamefont {A.}~\bibnamefont {Boatwright}}, \bibinfo {author} {\bibfnamefont {G.}~\bibnamefont {Sarma}}, \bibinfo {author} {\bibfnamefont {A.~M.}\ \bibnamefont {Sadoon}}, \bibinfo {author} {\bibfnamefont {A.}~\bibnamefont {Shirley}}, \bibinfo {author} {\bibfnamefont {N.}~\bibnamefont {Das Neves~Rodrigues}}, \bibinfo {author} {\bibfnamefont {E.~M.}\ \bibnamefont {Cunningham}}, \bibinfo {author} {\bibfnamefont {S.}~\bibnamefont {Yang}},\ and\ \bibinfo {author} {\bibfnamefont {A.~M.}\ \bibnamefont {Ellis}},\ }\bibfield  {title} {\bibinfo {title} {Communication: Infrared spectroscopy of salt-water complexes},\ }\href@noop {} {\bibfield  {journal} {\bibinfo  {journal} {J. Chem. Phys.}\ }\textbf {\bibinfo {volume} {144}} (\bibinfo {year} {2016})}\BibitemShut {NoStop}%
\bibitem [{\citenamefont {Gregoire}\ \emph {et~al.}(1998)\citenamefont {Gregoire}, \citenamefont {Mons}, \citenamefont {Dedonder-Lardeux},\ and\ \citenamefont {Jouvet}}]{gregoire1998nai}%
  \BibitemOpen
  \bibfield  {author} {\bibinfo {author} {\bibfnamefont {G.}~\bibnamefont {Gregoire}}, \bibinfo {author} {\bibfnamefont {M.}~\bibnamefont {Mons}}, \bibinfo {author} {\bibfnamefont {C.}~\bibnamefont {Dedonder-Lardeux}},\ and\ \bibinfo {author} {\bibfnamefont {C.}~\bibnamefont {Jouvet}},\ }\bibfield  {title} {\bibinfo {title} {Is \mbox{NaI} soluble in water clusters?},\ }\href@noop {} {\bibfield  {journal} {\bibinfo  {journal} {Eur. Phys. J. D}\ }\textbf {\bibinfo {volume} {1}},\ \bibinfo {pages} {5} (\bibinfo {year} {1998})}\BibitemShut {NoStop}%
\bibitem [{\citenamefont {M\"uller}\ \emph {et~al.}(2009)\citenamefont {M\"uller}, \citenamefont {Krapf}, \citenamefont {Koslowski}, \citenamefont {Mudrich},\ and\ \citenamefont {Stienkemeier}}]{MuellerPRL:2009}%
  \BibitemOpen
  \bibfield  {author} {\bibinfo {author} {\bibfnamefont {S.}~\bibnamefont {M\"uller}}, \bibinfo {author} {\bibfnamefont {S.}~\bibnamefont {Krapf}}, \bibinfo {author} {\bibfnamefont {T.}~\bibnamefont {Koslowski}}, \bibinfo {author} {\bibfnamefont {M.}~\bibnamefont {Mudrich}},\ and\ \bibinfo {author} {\bibfnamefont {F.}~\bibnamefont {Stienkemeier}},\ }\bibfield  {title} {\bibinfo {title} {Cold reactions of alkali-metal and water clusters inside helium nanodroplets},\ }\href@noop {} {\bibfield  {journal} {\bibinfo  {journal} {Phys. Rev. Lett.}\ }\textbf {\bibinfo {volume} {102}},\ \bibinfo {pages} {183401} (\bibinfo {year} {2009})}\BibitemShut {NoStop}%
\bibitem [{\citenamefont {Krasnokutski}\ \emph {et~al.}(2016)\citenamefont {Krasnokutski}, \citenamefont {Kuhn}, \citenamefont {Renzler}, \citenamefont {J{\"a}ger}, \citenamefont {Henning},\ and\ \citenamefont {Scheier}}]{krasnokutski2016ultra}%
  \BibitemOpen
  \bibfield  {author} {\bibinfo {author} {\bibfnamefont {S.}~\bibnamefont {Krasnokutski}}, \bibinfo {author} {\bibfnamefont {M.}~\bibnamefont {Kuhn}}, \bibinfo {author} {\bibfnamefont {M.}~\bibnamefont {Renzler}}, \bibinfo {author} {\bibfnamefont {C.}~\bibnamefont {J{\"a}ger}}, \bibinfo {author} {\bibfnamefont {T.}~\bibnamefont {Henning}},\ and\ \bibinfo {author} {\bibfnamefont {P.}~\bibnamefont {Scheier}},\ }\bibfield  {title} {\bibinfo {title} {Ultra-low-temperature reactions of carbon atoms with hydrogen molecules},\ }\href@noop {} {\bibfield  {journal} {\bibinfo  {journal} {Astrophys. J. Lett.}\ }\textbf {\bibinfo {volume} {818}},\ \bibinfo {pages} {L31} (\bibinfo {year} {2016})}\BibitemShut {NoStop}%
\bibitem [{\citenamefont {Mani}\ \emph {et~al.}(2019)\citenamefont {Mani}, \citenamefont {de~Tudela}, \citenamefont {Schwan}, \citenamefont {Pal}, \citenamefont {K{\"o}rning}, \citenamefont {Forbert}, \citenamefont {Redlich}, \citenamefont {Van Der~Meer}, \citenamefont {Schwaab}, \citenamefont {Marx} \emph {et~al.}}]{mani2019acid}%
  \BibitemOpen
  \bibfield  {author} {\bibinfo {author} {\bibfnamefont {D.}~\bibnamefont {Mani}}, \bibinfo {author} {\bibfnamefont {R.~P.}\ \bibnamefont {de~Tudela}}, \bibinfo {author} {\bibfnamefont {R.}~\bibnamefont {Schwan}}, \bibinfo {author} {\bibfnamefont {N.}~\bibnamefont {Pal}}, \bibinfo {author} {\bibfnamefont {S.}~\bibnamefont {K{\"o}rning}}, \bibinfo {author} {\bibfnamefont {H.}~\bibnamefont {Forbert}}, \bibinfo {author} {\bibfnamefont {B.}~\bibnamefont {Redlich}}, \bibinfo {author} {\bibfnamefont {A.}~\bibnamefont {Van Der~Meer}}, \bibinfo {author} {\bibfnamefont {G.}~\bibnamefont {Schwaab}}, \bibinfo {author} {\bibfnamefont {D.}~\bibnamefont {Marx}}, \emph {et~al.},\ }\bibfield  {title} {\bibinfo {title} {Acid solvation versus dissociation at “stardust conditions”: Reaction sequence matters},\ }\href@noop {} {\bibfield  {journal} {\bibinfo  {journal} {Sci. Adv.}\ }\textbf {\bibinfo {volume} {5}},\ \bibinfo {pages} {eaav8179} (\bibinfo {year} {2019})}\BibitemShut {NoStop}%
\bibitem [{\citenamefont {Albrechtsen}\ \emph {et~al.}(2023)\citenamefont {Albrechtsen}, \citenamefont {Schouder}, \citenamefont {Vi{\~n}as~Mu{\~n}oz}, \citenamefont {Christensen}, \citenamefont {Engelbrecht~Petersen}, \citenamefont {Pi}, \citenamefont {Barranco},\ and\ \citenamefont {Stapelfeldt}}]{albrechtsen2023observing}%
  \BibitemOpen
  \bibfield  {author} {\bibinfo {author} {\bibfnamefont {S.~H.}\ \bibnamefont {Albrechtsen}}, \bibinfo {author} {\bibfnamefont {C.~A.}\ \bibnamefont {Schouder}}, \bibinfo {author} {\bibfnamefont {A.}~\bibnamefont {Vi{\~n}as~Mu{\~n}oz}}, \bibinfo {author} {\bibfnamefont {J.~K.}\ \bibnamefont {Christensen}}, \bibinfo {author} {\bibfnamefont {C.}~\bibnamefont {Engelbrecht~Petersen}}, \bibinfo {author} {\bibfnamefont {M.}~\bibnamefont {Pi}}, \bibinfo {author} {\bibfnamefont {M.}~\bibnamefont {Barranco}},\ and\ \bibinfo {author} {\bibfnamefont {H.}~\bibnamefont {Stapelfeldt}},\ }\bibfield  {title} {\bibinfo {title} {Observing the primary steps of ion solvation in helium droplets},\ }\href@noop {} {\bibfield  {journal} {\bibinfo  {journal} {Nature}\ }\textbf {\bibinfo {volume} {623}},\ \bibinfo {pages} {319} (\bibinfo {year} {2023})}\BibitemShut {NoStop}%
\bibitem [{\citenamefont {Nauta}\ and\ \citenamefont {Miller}(2000)}]{nauta2000formation}%
  \BibitemOpen
  \bibfield  {author} {\bibinfo {author} {\bibfnamefont {K.}~\bibnamefont {Nauta}}\ and\ \bibinfo {author} {\bibfnamefont {R.}~\bibnamefont {Miller}},\ }\bibfield  {title} {\bibinfo {title} {Formation of cyclic water hexamer in liquid helium: The smallest piece of ice},\ }\href@noop {} {\bibfield  {journal} {\bibinfo  {journal} {Science}\ }\textbf {\bibinfo {volume} {287}},\ \bibinfo {pages} {293} (\bibinfo {year} {2000})}\BibitemShut {NoStop}%
\bibitem [{\citenamefont {Gutberlet}\ \emph {et~al.}(2009)\citenamefont {Gutberlet}, \citenamefont {Schwaab}, \citenamefont {Birer}, \citenamefont {Masia}, \citenamefont {Kaczmarek}, \citenamefont {Forbert}, \citenamefont {Havenith},\ and\ \citenamefont {Marx}}]{gutberlet2009aggregation}%
  \BibitemOpen
  \bibfield  {author} {\bibinfo {author} {\bibfnamefont {A.}~\bibnamefont {Gutberlet}}, \bibinfo {author} {\bibfnamefont {G.}~\bibnamefont {Schwaab}}, \bibinfo {author} {\bibfnamefont {{\"O}.}~\bibnamefont {Birer}}, \bibinfo {author} {\bibfnamefont {M.}~\bibnamefont {Masia}}, \bibinfo {author} {\bibfnamefont {A.}~\bibnamefont {Kaczmarek}}, \bibinfo {author} {\bibfnamefont {H.}~\bibnamefont {Forbert}}, \bibinfo {author} {\bibfnamefont {M.}~\bibnamefont {Havenith}},\ and\ \bibinfo {author} {\bibfnamefont {D.}~\bibnamefont {Marx}},\ }\bibfield  {title} {\bibinfo {title} {Aggregation-induced dissociation of \mbox{HCl(H$_2$O)}$_4$ below 1~\uppercase{K}: The smallest droplet of acid},\ }\href@noop {} {\bibfield  {journal} {\bibinfo  {journal} {Science}\ }\textbf {\bibinfo {volume} {324}},\ \bibinfo {pages} {1545} (\bibinfo {year} {2009})}\BibitemShut {NoStop}%
\bibitem [{\citenamefont {Ale{\v{s}}kovi{\'c}}\ \emph {et~al.}(2023)\citenamefont {Ale{\v{s}}kovi{\'c}}, \citenamefont {K{\"u}stner}, \citenamefont {Messner}, \citenamefont {Lackner}, \citenamefont {Ernst},\ and\ \citenamefont {{\v{S}}ekutor}}]{alevskovic2023nanostructured}%
  \BibitemOpen
  \bibfield  {author} {\bibinfo {author} {\bibfnamefont {M.}~\bibnamefont {Ale{\v{s}}kovi{\'c}}}, \bibinfo {author} {\bibfnamefont {F.}~\bibnamefont {K{\"u}stner}}, \bibinfo {author} {\bibfnamefont {R.}~\bibnamefont {Messner}}, \bibinfo {author} {\bibfnamefont {F.}~\bibnamefont {Lackner}}, \bibinfo {author} {\bibfnamefont {W.~E.}\ \bibnamefont {Ernst}},\ and\ \bibinfo {author} {\bibfnamefont {M.}~\bibnamefont {{\v{S}}ekutor}},\ }\bibfield  {title} {\bibinfo {title} {Nanostructured supramolecular networks from self-assembled diamondoid molecules under ultracold conditions},\ }\href@noop {} {\bibfield  {journal} {\bibinfo  {journal} {Phys. Chem. Chem. Phys.}\ }\textbf {\bibinfo {volume} {25}},\ \bibinfo {pages} {17869} (\bibinfo {year} {2023})}\BibitemShut {NoStop}%
\bibitem [{\citenamefont {Buchta}\ \emph {et~al.}(2013{\natexlab{a}})\citenamefont {Buchta}, \citenamefont {Krishnan}, \citenamefont {Brauer}, \citenamefont {Drabbels}, \citenamefont {O'Keeffe}, \citenamefont {Devetta}, \citenamefont {Di~Fraia}, \citenamefont {Callegari}, \citenamefont {Richter}, \citenamefont {Coreno}, \citenamefont {Prince}, \citenamefont {Stienkemeier}, \citenamefont {Moshammer},\ and\ \citenamefont {Mudrich}}]{Buchta:2013}%
  \BibitemOpen
  \bibfield  {author} {\bibinfo {author} {\bibfnamefont {D.}~\bibnamefont {Buchta}}, \bibinfo {author} {\bibfnamefont {S.~R.}\ \bibnamefont {Krishnan}}, \bibinfo {author} {\bibfnamefont {N.~B.}\ \bibnamefont {Brauer}}, \bibinfo {author} {\bibfnamefont {M.}~\bibnamefont {Drabbels}}, \bibinfo {author} {\bibfnamefont {P.}~\bibnamefont {O'Keeffe}}, \bibinfo {author} {\bibfnamefont {M.}~\bibnamefont {Devetta}}, \bibinfo {author} {\bibfnamefont {M.}~\bibnamefont {Di~Fraia}}, \bibinfo {author} {\bibfnamefont {C.}~\bibnamefont {Callegari}}, \bibinfo {author} {\bibfnamefont {R.}~\bibnamefont {Richter}}, \bibinfo {author} {\bibfnamefont {M.}~\bibnamefont {Coreno}}, \bibinfo {author} {\bibfnamefont {K.~C.}\ \bibnamefont {Prince}}, \bibinfo {author} {\bibfnamefont {F.}~\bibnamefont {Stienkemeier}}, \bibinfo {author} {\bibfnamefont {R.}~\bibnamefont {Moshammer}},\ and\ \bibinfo {author} {\bibfnamefont {M.}~\bibnamefont {Mudrich}},\ }\bibfield  {title} {\bibinfo {title} {Charge transfer and penning ionization of dopants in
  or on helium nanodroplets exposed to {EUV} radiation},\ }\href {https://doi.org/10.1021/jp401424w} {\bibfield  {journal} {\bibinfo  {journal} {J. Phys. Chem. A}\ }\textbf {\bibinfo {volume} {117}},\ \bibinfo {pages} {4394} (\bibinfo {year} {2013}{\natexlab{a}})}\BibitemShut {NoStop}%
\bibitem [{\citenamefont {Buchta}\ \emph {et~al.}(2013{\natexlab{b}})\citenamefont {Buchta}, \citenamefont {Krishnan}, \citenamefont {Brauer}, \citenamefont {Drabbels}, \citenamefont {O'Keeffe}, \citenamefont {Devetta}, \citenamefont {Di~Fraia}, \citenamefont {Callegari}, \citenamefont {Richter}, \citenamefont {Coreno}, \citenamefont {Prince}, \citenamefont {Stienkemeier}, \citenamefont {Ullrich}, \citenamefont {Moshammer},\ and\ \citenamefont {Mudrich}}]{BuchtaJCP:2013}%
  \BibitemOpen
  \bibfield  {author} {\bibinfo {author} {\bibfnamefont {D.}~\bibnamefont {Buchta}}, \bibinfo {author} {\bibfnamefont {S.~R.}\ \bibnamefont {Krishnan}}, \bibinfo {author} {\bibfnamefont {N.~B.}\ \bibnamefont {Brauer}}, \bibinfo {author} {\bibfnamefont {M.}~\bibnamefont {Drabbels}}, \bibinfo {author} {\bibfnamefont {P.}~\bibnamefont {O'Keeffe}}, \bibinfo {author} {\bibfnamefont {M.}~\bibnamefont {Devetta}}, \bibinfo {author} {\bibfnamefont {M.}~\bibnamefont {Di~Fraia}}, \bibinfo {author} {\bibfnamefont {C.}~\bibnamefont {Callegari}}, \bibinfo {author} {\bibfnamefont {R.}~\bibnamefont {Richter}}, \bibinfo {author} {\bibfnamefont {M.}~\bibnamefont {Coreno}}, \bibinfo {author} {\bibfnamefont {K.~C.}\ \bibnamefont {Prince}}, \bibinfo {author} {\bibfnamefont {F.}~\bibnamefont {Stienkemeier}}, \bibinfo {author} {\bibfnamefont {J.}~\bibnamefont {Ullrich}}, \bibinfo {author} {\bibfnamefont {R.}~\bibnamefont {Moshammer}},\ and\ \bibinfo {author} {\bibfnamefont {M.}~\bibnamefont {Mudrich}},\ }\bibfield  {title}
  {\bibinfo {title} {Extreme ultraviolet ionization of pure {He} nanodroplets: Mass-correlated photoelectron imaging, penning ionization, and electron energy-loss spectra},\ }\href@noop {} {\bibfield  {journal} {\bibinfo  {journal} {J. Chem. Phys.}\ }\textbf {\bibinfo {volume} {139}},\ \bibinfo {pages} {084301} (\bibinfo {year} {2013}{\natexlab{b}})}\BibitemShut {NoStop}%
\bibitem [{\citenamefont {LaForge}\ \emph {et~al.}(2024)\citenamefont {LaForge}, \citenamefont {Ltaief}, \citenamefont {Krishnan}, \citenamefont {Sisourat},\ and\ \citenamefont {Mudrich}}]{laforge2024interatomic}%
  \BibitemOpen
  \bibfield  {author} {\bibinfo {author} {\bibfnamefont {A.~C.}\ \bibnamefont {LaForge}}, \bibinfo {author} {\bibfnamefont {L.~B.}\ \bibnamefont {Ltaief}}, \bibinfo {author} {\bibfnamefont {S.}~\bibnamefont {Krishnan}}, \bibinfo {author} {\bibfnamefont {N.}~\bibnamefont {Sisourat}},\ and\ \bibinfo {author} {\bibfnamefont {M.}~\bibnamefont {Mudrich}},\ }\bibfield  {title} {\bibinfo {title} {Interatomic and intermolecular decay processes in quantum fluid clusters},\ }\href@noop {} {\bibfield  {journal} {\bibinfo  {journal} {Rep. Prog. Phys.}\ }\textbf {\bibinfo {volume} {87}},\ \bibinfo {pages} {126402} (\bibinfo {year} {2024})}\BibitemShut {NoStop}%
\bibitem [{\citenamefont {LaForge}\ \emph {et~al.}(2016)\citenamefont {LaForge}, \citenamefont {Stumpf}, \citenamefont {Gokhberg}, \citenamefont {von Vangerow}, \citenamefont {Stienkemeier}, \citenamefont {Kryzhevoi}, \citenamefont {O'Keeffe}, \citenamefont {Ciavardini}, \citenamefont {Krishnan}, \citenamefont {Coreno}, \citenamefont {Prince}, \citenamefont {Richter}, \citenamefont {Moshammer}, \citenamefont {Pfeifer}, \citenamefont {Cederbaum},\ and\ \citenamefont {Mudrich}}]{LaForgePRL:2016}%
  \BibitemOpen
  \bibfield  {author} {\bibinfo {author} {\bibfnamefont {A.~C.}\ \bibnamefont {LaForge}}, \bibinfo {author} {\bibfnamefont {V.}~\bibnamefont {Stumpf}}, \bibinfo {author} {\bibfnamefont {K.}~\bibnamefont {Gokhberg}}, \bibinfo {author} {\bibfnamefont {J.}~\bibnamefont {von Vangerow}}, \bibinfo {author} {\bibfnamefont {F.}~\bibnamefont {Stienkemeier}}, \bibinfo {author} {\bibfnamefont {N.~V.}\ \bibnamefont {Kryzhevoi}}, \bibinfo {author} {\bibfnamefont {P.}~\bibnamefont {O'Keeffe}}, \bibinfo {author} {\bibfnamefont {A.}~\bibnamefont {Ciavardini}}, \bibinfo {author} {\bibfnamefont {S.~R.}\ \bibnamefont {Krishnan}}, \bibinfo {author} {\bibfnamefont {M.}~\bibnamefont {Coreno}}, \bibinfo {author} {\bibfnamefont {K.~C.}\ \bibnamefont {Prince}}, \bibinfo {author} {\bibfnamefont {R.}~\bibnamefont {Richter}}, \bibinfo {author} {\bibfnamefont {R.}~\bibnamefont {Moshammer}}, \bibinfo {author} {\bibfnamefont {T.}~\bibnamefont {Pfeifer}}, \bibinfo {author} {\bibfnamefont {L.~S.}\ \bibnamefont {Cederbaum}},\ and\ \bibinfo
  {author} {\bibfnamefont {M.}~\bibnamefont {Mudrich}},\ }\bibfield  {title} {\bibinfo {title} {Enhanced ionization of embedded clusters by electron-transfer-mediated decay in helium nanodroplets},\ }\href@noop {} {\bibfield  {journal} {\bibinfo  {journal} {Phys. Rev. Lett.}\ }\textbf {\bibinfo {volume} {116}},\ \bibinfo {pages} {203001} (\bibinfo {year} {2016})}\BibitemShut {NoStop}%
\bibitem [{\citenamefont {LaForge}\ \emph {et~al.}(2019)\citenamefont {LaForge}, \citenamefont {Shcherbinin}, \citenamefont {Stienkemeier}, \citenamefont {Richter}, \citenamefont {Moshammer}, \citenamefont {Pfeifer},\ and\ \citenamefont {Mudrich}}]{LaForge:2019}%
  \BibitemOpen
  \bibfield  {author} {\bibinfo {author} {\bibfnamefont {A.}~\bibnamefont {LaForge}}, \bibinfo {author} {\bibfnamefont {M.}~\bibnamefont {Shcherbinin}}, \bibinfo {author} {\bibfnamefont {F.}~\bibnamefont {Stienkemeier}}, \bibinfo {author} {\bibfnamefont {R.}~\bibnamefont {Richter}}, \bibinfo {author} {\bibfnamefont {R.}~\bibnamefont {Moshammer}}, \bibinfo {author} {\bibfnamefont {T.}~\bibnamefont {Pfeifer}},\ and\ \bibinfo {author} {\bibfnamefont {M.}~\bibnamefont {Mudrich}},\ }\bibfield  {title} {\bibinfo {title} {Highly efficient double ionization of mixed alkali dimers by intermolecular coulombic decay},\ }\href@noop {} {\bibfield  {journal} {\bibinfo  {journal} {Nat. Phys.}\ }\textbf {\bibinfo {volume} {15}},\ \bibinfo {pages} {247} (\bibinfo {year} {2019})}\BibitemShut {NoStop}%
\bibitem [{\citenamefont {Ben~Ltaief}\ \emph {et~al.}(2023)\citenamefont {Ben~Ltaief}, \citenamefont {Sishodia}, \citenamefont {Mandal}, \citenamefont {De}, \citenamefont {Krishnan}, \citenamefont {Medina}, \citenamefont {Pal}, \citenamefont {Richter}, \citenamefont {Fennel},\ and\ \citenamefont {Mudrich}}]{ltaief:2023}%
  \BibitemOpen
  \bibfield  {author} {\bibinfo {author} {\bibfnamefont {L.}~\bibnamefont {Ben~Ltaief}}, \bibinfo {author} {\bibfnamefont {K.}~\bibnamefont {Sishodia}}, \bibinfo {author} {\bibfnamefont {S.}~\bibnamefont {Mandal}}, \bibinfo {author} {\bibfnamefont {S.}~\bibnamefont {De}}, \bibinfo {author} {\bibfnamefont {S.~R.}\ \bibnamefont {Krishnan}}, \bibinfo {author} {\bibfnamefont {C.}~\bibnamefont {Medina}}, \bibinfo {author} {\bibfnamefont {N.}~\bibnamefont {Pal}}, \bibinfo {author} {\bibfnamefont {R.}~\bibnamefont {Richter}}, \bibinfo {author} {\bibfnamefont {T.}~\bibnamefont {Fennel}},\ and\ \bibinfo {author} {\bibfnamefont {M.}~\bibnamefont {Mudrich}},\ }\bibfield  {title} {\bibinfo {title} {Efficient indirect interatomic coulombic decay induced by photoelectron impact excitation in large pure helium nanodroplets},\ }\href@noop {} {\bibfield  {journal} {\bibinfo  {journal} {Phys. Rev. Lett.}\ }\textbf {\bibinfo {volume} {131}},\ \bibinfo {pages} {023001} (\bibinfo {year} {2023})}\BibitemShut {NoStop}%
\bibitem [{\citenamefont {Ben~Ltaief}\ \emph {et~al.}(2024)\citenamefont {Ben~Ltaief}, \citenamefont {Sishodia}, \citenamefont {Richter}, \citenamefont {Bastian}, \citenamefont {Asmussen}, \citenamefont {Mandal}, \citenamefont {Pal}, \citenamefont {Medina}, \citenamefont {Krishnan}, \citenamefont {von Haeften},\ and\ \citenamefont {Mudrich}}]{ltaiefPRR:2024}%
  \BibitemOpen
  \bibfield  {author} {\bibinfo {author} {\bibfnamefont {L.}~\bibnamefont {Ben~Ltaief}}, \bibinfo {author} {\bibfnamefont {K.}~\bibnamefont {Sishodia}}, \bibinfo {author} {\bibfnamefont {R.}~\bibnamefont {Richter}}, \bibinfo {author} {\bibfnamefont {B.}~\bibnamefont {Bastian}}, \bibinfo {author} {\bibfnamefont {J.~D.}\ \bibnamefont {Asmussen}}, \bibinfo {author} {\bibfnamefont {S.}~\bibnamefont {Mandal}}, \bibinfo {author} {\bibfnamefont {N.}~\bibnamefont {Pal}}, \bibinfo {author} {\bibfnamefont {C.}~\bibnamefont {Medina}}, \bibinfo {author} {\bibfnamefont {S.~R.}\ \bibnamefont {Krishnan}}, \bibinfo {author} {\bibfnamefont {K.}~\bibnamefont {von Haeften}},\ and\ \bibinfo {author} {\bibfnamefont {M.}~\bibnamefont {Mudrich}},\ }\bibfield  {title} {\bibinfo {title} {Spectroscopically resolved resonant interatomic coulombic decay in photoexcited large \mbox{He} nanodroplets},\ }\href@noop {} {\bibfield  {journal} {\bibinfo  {journal} {Phys. Rev. Res.}\ }\textbf {\bibinfo {volume} {6}},\ \bibinfo {pages} {013019}
  (\bibinfo {year} {2024})}\BibitemShut {NoStop}%
\bibitem [{\citenamefont {Bastian}\ \emph {et~al.}(2024)\citenamefont {Bastian}, \citenamefont {Asmussen}, \citenamefont {Ltaief}, \citenamefont {Pedersen}, \citenamefont {Sishodia}, \citenamefont {De}, \citenamefont {Krishnan}, \citenamefont {Medina}, \citenamefont {Pal}, \citenamefont {Richter} \emph {et~al.}}]{bastian2024observation}%
  \BibitemOpen
  \bibfield  {author} {\bibinfo {author} {\bibfnamefont {B.}~\bibnamefont {Bastian}}, \bibinfo {author} {\bibfnamefont {J.}~\bibnamefont {Asmussen}}, \bibinfo {author} {\bibfnamefont {L.~B.}\ \bibnamefont {Ltaief}}, \bibinfo {author} {\bibfnamefont {H.}~\bibnamefont {Pedersen}}, \bibinfo {author} {\bibfnamefont {K.}~\bibnamefont {Sishodia}}, \bibinfo {author} {\bibfnamefont {S.}~\bibnamefont {De}}, \bibinfo {author} {\bibfnamefont {S.}~\bibnamefont {Krishnan}}, \bibinfo {author} {\bibfnamefont {C.}~\bibnamefont {Medina}}, \bibinfo {author} {\bibfnamefont {N.}~\bibnamefont {Pal}}, \bibinfo {author} {\bibfnamefont {R.}~\bibnamefont {Richter}}, \emph {et~al.},\ }\bibfield  {title} {\bibinfo {title} {Observation of interatomic coulombic decay induced by double excitation of helium in nanodroplets},\ }\href@noop {} {\bibfield  {journal} {\bibinfo  {journal} {Phys. Rev. Lett.}\ }\textbf {\bibinfo {volume} {132}},\ \bibinfo {pages} {233001} (\bibinfo {year} {2024})}\BibitemShut {NoStop}%
\bibitem [{\citenamefont {Ben~Ltaief}\ \emph {et~al.}(2019)\citenamefont {Ben~Ltaief}, \citenamefont {Shcherbinin}, \citenamefont {Mandal}, \citenamefont {Krishnan}, \citenamefont {LaForge}, \citenamefont {Richter}, \citenamefont {Turchini}, \citenamefont {Zema}, \citenamefont {Pfeifer}, \citenamefont {Fasshauer} \emph {et~al.}}]{Ltaief:2019}%
  \BibitemOpen
  \bibfield  {author} {\bibinfo {author} {\bibfnamefont {L.}~\bibnamefont {Ben~Ltaief}}, \bibinfo {author} {\bibfnamefont {M.}~\bibnamefont {Shcherbinin}}, \bibinfo {author} {\bibfnamefont {S.}~\bibnamefont {Mandal}}, \bibinfo {author} {\bibfnamefont {S.}~\bibnamefont {Krishnan}}, \bibinfo {author} {\bibfnamefont {A.}~\bibnamefont {LaForge}}, \bibinfo {author} {\bibfnamefont {R.}~\bibnamefont {Richter}}, \bibinfo {author} {\bibfnamefont {S.}~\bibnamefont {Turchini}}, \bibinfo {author} {\bibfnamefont {N.}~\bibnamefont {Zema}}, \bibinfo {author} {\bibfnamefont {T.}~\bibnamefont {Pfeifer}}, \bibinfo {author} {\bibfnamefont {E.}~\bibnamefont {Fasshauer}}, \emph {et~al.},\ }\bibfield  {title} {\bibinfo {title} {Charge exchange dominates long-range interatomic coulombic decay of excited metal-doped helium nanodroplets},\ }\href@noop {} {\bibfield  {journal} {\bibinfo  {journal} {J. Phys. Chem. Lett.}\ }\textbf {\bibinfo {volume} {10}},\ \bibinfo {pages} {6904} (\bibinfo {year} {2019})}\BibitemShut {NoStop}%
\bibitem [{\citenamefont {Shcherbinin}\ \emph {et~al.}(2018)\citenamefont {Shcherbinin}, \citenamefont {LaForge}, \citenamefont {Hanif}, \citenamefont {Richter},\ and\ \citenamefont {Mudrich}}]{Shcherbinin:2018}%
  \BibitemOpen
  \bibfield  {author} {\bibinfo {author} {\bibfnamefont {M.}~\bibnamefont {Shcherbinin}}, \bibinfo {author} {\bibfnamefont {A.~C.}\ \bibnamefont {LaForge}}, \bibinfo {author} {\bibfnamefont {M.}~\bibnamefont {Hanif}}, \bibinfo {author} {\bibfnamefont {R.}~\bibnamefont {Richter}},\ and\ \bibinfo {author} {\bibfnamefont {M.}~\bibnamefont {Mudrich}},\ }\bibfield  {title} {\bibinfo {title} {Penning ionization of acene molecules by helium nanodroplets},\ }\href@noop {} {\bibfield  {journal} {\bibinfo  {journal} {J. Phys. Chem. A}\ }\textbf {\bibinfo {volume} {122}},\ \bibinfo {pages} {1855} (\bibinfo {year} {2018})}\BibitemShut {NoStop}%
\bibitem [{\citenamefont {Ltaief}\ \emph {et~al.}(2020)\citenamefont {Ltaief}, \citenamefont {Shcherbinin}, \citenamefont {Krishnan}, \citenamefont {Richter}, \citenamefont {Pfeifer}, \citenamefont {Bauer}, \citenamefont {Ghosh}, \citenamefont {Mudrich}, \citenamefont {Gokhberg}, \citenamefont {Laforge} \emph {et~al.}}]{Ltaief:2020}%
  \BibitemOpen
  \bibfield  {author} {\bibinfo {author} {\bibfnamefont {L.~B.}\ \bibnamefont {Ltaief}}, \bibinfo {author} {\bibfnamefont {M.}~\bibnamefont {Shcherbinin}}, \bibinfo {author} {\bibfnamefont {S.}~\bibnamefont {Krishnan}}, \bibinfo {author} {\bibfnamefont {R.}~\bibnamefont {Richter}}, \bibinfo {author} {\bibfnamefont {T.}~\bibnamefont {Pfeifer}}, \bibinfo {author} {\bibfnamefont {M.}~\bibnamefont {Bauer}}, \bibinfo {author} {\bibfnamefont {A.}~\bibnamefont {Ghosh}}, \bibinfo {author} {\bibfnamefont {M.}~\bibnamefont {Mudrich}}, \bibinfo {author} {\bibfnamefont {K.}~\bibnamefont {Gokhberg}}, \bibinfo {author} {\bibfnamefont {A.}~\bibnamefont {Laforge}}, \emph {et~al.},\ }\bibfield  {title} {\bibinfo {title} {Electron transfer mediated decay of alkali dimers attached to {He} nanodroplets},\ }\href@noop {} {\bibfield  {journal} {\bibinfo  {journal} {Phys. Chem. Chem. Phys.}\ }\textbf {\bibinfo {volume} {22}},\ \bibinfo {pages} {8557} (\bibinfo {year} {2020})}\BibitemShut {NoStop}%
\bibitem [{\citenamefont {Mandal}\ \emph {et~al.}(2020)\citenamefont {Mandal}, \citenamefont {Gopal}, \citenamefont {Shcherbinin}, \citenamefont {D’Elia}, \citenamefont {Srinivas}, \citenamefont {Richter}, \citenamefont {Coreno}, \citenamefont {Bapat}, \citenamefont {Mudrich}, \citenamefont {Krishnan} \emph {et~al.}}]{Mandal2020}%
  \BibitemOpen
  \bibfield  {author} {\bibinfo {author} {\bibfnamefont {S.}~\bibnamefont {Mandal}}, \bibinfo {author} {\bibfnamefont {R.}~\bibnamefont {Gopal}}, \bibinfo {author} {\bibfnamefont {M.}~\bibnamefont {Shcherbinin}}, \bibinfo {author} {\bibfnamefont {A.}~\bibnamefont {D’Elia}}, \bibinfo {author} {\bibfnamefont {H.}~\bibnamefont {Srinivas}}, \bibinfo {author} {\bibfnamefont {R.}~\bibnamefont {Richter}}, \bibinfo {author} {\bibfnamefont {M.}~\bibnamefont {Coreno}}, \bibinfo {author} {\bibfnamefont {B.}~\bibnamefont {Bapat}}, \bibinfo {author} {\bibfnamefont {M.}~\bibnamefont {Mudrich}}, \bibinfo {author} {\bibfnamefont {S.~R.}\ \bibnamefont {Krishnan}}, \emph {et~al.},\ }\bibfield  {title} {\bibinfo {title} {Penning spectroscopy and structure of acetylene oligomers in he nanodroplets},\ }\href@noop {} {\bibfield  {journal} {\bibinfo  {journal} {Phys. Chem. Chem. Phys.}\ }\textbf {\bibinfo {volume} {22}},\ \bibinfo {pages} {10149} (\bibinfo {year} {2020})}\BibitemShut {NoStop}%
\bibitem [{\citenamefont {Asmussen}\ \emph {et~al.}(2023{\natexlab{a}})\citenamefont {Asmussen}, \citenamefont {Abid}, \citenamefont {Sundaralingam}, \citenamefont {Bastian}, \citenamefont {Sishodia}, \citenamefont {De}, \citenamefont {Ltaief}, \citenamefont {Krishnan}, \citenamefont {Pedersen},\ and\ \citenamefont {Mudrich}}]{asmussen2023secondary}%
  \BibitemOpen
  \bibfield  {author} {\bibinfo {author} {\bibfnamefont {J.~D.}\ \bibnamefont {Asmussen}}, \bibinfo {author} {\bibfnamefont {A.~R.}\ \bibnamefont {Abid}}, \bibinfo {author} {\bibfnamefont {A.}~\bibnamefont {Sundaralingam}}, \bibinfo {author} {\bibfnamefont {B.}~\bibnamefont {Bastian}}, \bibinfo {author} {\bibfnamefont {K.}~\bibnamefont {Sishodia}}, \bibinfo {author} {\bibfnamefont {S.}~\bibnamefont {De}}, \bibinfo {author} {\bibfnamefont {L.~B.}\ \bibnamefont {Ltaief}}, \bibinfo {author} {\bibfnamefont {S.}~\bibnamefont {Krishnan}}, \bibinfo {author} {\bibfnamefont {H.~B.}\ \bibnamefont {Pedersen}},\ and\ \bibinfo {author} {\bibfnamefont {M.}~\bibnamefont {Mudrich}},\ }\bibfield  {title} {\bibinfo {title} {Secondary ionization of pyrimidine nucleobases and their microhydrated derivatives in helium nanodroplets},\ }\href@noop {} {\bibfield  {journal} {\bibinfo  {journal} {Phys. Chem. Chem. Phys.}\ }\textbf {\bibinfo {volume} {25}},\ \bibinfo {pages} {24819} (\bibinfo {year} {2023}{\natexlab{a}})}\BibitemShut
  {NoStop}%
\bibitem [{\citenamefont {Sen}\ \emph {et~al.}(2024)\citenamefont {Sen}, \citenamefont {Mandal}, \citenamefont {De}, \citenamefont {Sen}, \citenamefont {Gopal}, \citenamefont {Ben~Ltaief}, \citenamefont {Turchini}, \citenamefont {Catone}, \citenamefont {Zema}, \citenamefont {Coreno} \emph {et~al.}}]{sen2024electron}%
  \BibitemOpen
  \bibfield  {author} {\bibinfo {author} {\bibfnamefont {S.}~\bibnamefont {Sen}}, \bibinfo {author} {\bibfnamefont {S.}~\bibnamefont {Mandal}}, \bibinfo {author} {\bibfnamefont {S.}~\bibnamefont {De}}, \bibinfo {author} {\bibfnamefont {A.}~\bibnamefont {Sen}}, \bibinfo {author} {\bibfnamefont {R.}~\bibnamefont {Gopal}}, \bibinfo {author} {\bibfnamefont {L.}~\bibnamefont {Ben~Ltaief}}, \bibinfo {author} {\bibfnamefont {S.}~\bibnamefont {Turchini}}, \bibinfo {author} {\bibfnamefont {D.}~\bibnamefont {Catone}}, \bibinfo {author} {\bibfnamefont {N.}~\bibnamefont {Zema}}, \bibinfo {author} {\bibfnamefont {M.}~\bibnamefont {Coreno}}, \emph {et~al.},\ }\bibfield  {title} {\bibinfo {title} {Electron and ion spectroscopy of camphor doped helium nanodroplets in the extreme \mbox{UV} and soft x-ray regime},\ }\href@noop {} {\bibfield  {journal} {\bibinfo  {journal} {J. Phys. B: At., Mol. Opt. Phys.}\ }\textbf {\bibinfo {volume} {57}},\ \bibinfo {pages} {015201} (\bibinfo {year} {2024})}\BibitemShut {NoStop}%
\bibitem [{\citenamefont {Ltaief}\ \emph {et~al.}(2025)\citenamefont {Ltaief}, \citenamefont {Sishodia}, \citenamefont {Asmussen}, \citenamefont {Abid}, \citenamefont {Krishnan}, \citenamefont {Pedersen}, \citenamefont {Sisourat},\ and\ \citenamefont {Mudrich}}]{ltaief2025interatomic}%
  \BibitemOpen
  \bibfield  {author} {\bibinfo {author} {\bibfnamefont {L.~B.}\ \bibnamefont {Ltaief}}, \bibinfo {author} {\bibfnamefont {K.}~\bibnamefont {Sishodia}}, \bibinfo {author} {\bibfnamefont {J.~D.}\ \bibnamefont {Asmussen}}, \bibinfo {author} {\bibfnamefont {A.~R.}\ \bibnamefont {Abid}}, \bibinfo {author} {\bibfnamefont {S.}~\bibnamefont {Krishnan}}, \bibinfo {author} {\bibfnamefont {H.~B.}\ \bibnamefont {Pedersen}}, \bibinfo {author} {\bibfnamefont {N.}~\bibnamefont {Sisourat}},\ and\ \bibinfo {author} {\bibfnamefont {M.}~\bibnamefont {Mudrich}},\ }\bibfield  {title} {\bibinfo {title} {Interatomic coulombic decay in lithium-doped large helium nanodroplets induced by photoelectron impact excitation},\ }\href@noop {} {\bibfield  {journal} {\bibinfo  {journal} {Rep. Prog. Phys.}\ }\textbf {\bibinfo {volume} {88}},\ \bibinfo {pages} {037901} (\bibinfo {year} {2025})}\BibitemShut {NoStop}%
\bibitem [{\citenamefont {Dupuy}\ \emph {et~al.}(2024)\citenamefont {Dupuy}, \citenamefont {Buttersack}, \citenamefont {Trinter}, \citenamefont {Richter}, \citenamefont {Gholami}, \citenamefont {Bj{\"o}rneholm}, \citenamefont {Hergenhahn}, \citenamefont {Winter},\ and\ \citenamefont {Bluhm}}]{dupuy:2024}%
  \BibitemOpen
  \bibfield  {author} {\bibinfo {author} {\bibfnamefont {R.}~\bibnamefont {Dupuy}}, \bibinfo {author} {\bibfnamefont {T.}~\bibnamefont {Buttersack}}, \bibinfo {author} {\bibfnamefont {F.}~\bibnamefont {Trinter}}, \bibinfo {author} {\bibfnamefont {C.}~\bibnamefont {Richter}}, \bibinfo {author} {\bibfnamefont {S.}~\bibnamefont {Gholami}}, \bibinfo {author} {\bibfnamefont {O.}~\bibnamefont {Bj{\"o}rneholm}}, \bibinfo {author} {\bibfnamefont {U.}~\bibnamefont {Hergenhahn}}, \bibinfo {author} {\bibfnamefont {B.}~\bibnamefont {Winter}},\ and\ \bibinfo {author} {\bibfnamefont {H.}~\bibnamefont {Bluhm}},\ }\bibfield  {title} {\bibinfo {title} {The solvation shell probed by resonant intermolecular coulombic decay},\ }\href@noop {} {\bibfield  {journal} {\bibinfo  {journal} {Nat. Commun.}\ }\textbf {\bibinfo {volume} {15}},\ \bibinfo {pages} {6926} (\bibinfo {year} {2024})}\BibitemShut {NoStop}%
\bibitem [{\citenamefont {Toennies}\ and\ \citenamefont {Vilesov}(2004)}]{Toennies:2004}%
  \BibitemOpen
  \bibfield  {author} {\bibinfo {author} {\bibfnamefont {J.~P.}\ \bibnamefont {Toennies}}\ and\ \bibinfo {author} {\bibfnamefont {A.~F.}\ \bibnamefont {Vilesov}},\ }\bibfield  {title} {\bibinfo {title} {Superfluid helium droplets: A uniquely cold nanomatrix for molecules and molecular complexes},\ }\href@noop {} {\bibfield  {journal} {\bibinfo  {journal} {Angew. Chem., Int. Ed. Engl.}\ }\textbf {\bibinfo {volume} {43}},\ \bibinfo {pages} {2622} (\bibinfo {year} {2004})}\BibitemShut {NoStop}%
\bibitem [{\citenamefont {Mudrich}\ \emph {et~al.}(2020)\citenamefont {Mudrich}, \citenamefont {LaForge}, \citenamefont {Ciavardini}, \citenamefont {O'Keeffe}, \citenamefont {Callegari}, \citenamefont {Coreno}, \citenamefont {Demidovich}, \citenamefont {Devetta}, \citenamefont {Di~Fraia}, \citenamefont {Drabbels} \emph {et~al.}}]{Mudrich:2020}%
  \BibitemOpen
  \bibfield  {author} {\bibinfo {author} {\bibfnamefont {M.}~\bibnamefont {Mudrich}}, \bibinfo {author} {\bibfnamefont {A.}~\bibnamefont {LaForge}}, \bibinfo {author} {\bibfnamefont {A.}~\bibnamefont {Ciavardini}}, \bibinfo {author} {\bibfnamefont {P.}~\bibnamefont {O'Keeffe}}, \bibinfo {author} {\bibfnamefont {C.}~\bibnamefont {Callegari}}, \bibinfo {author} {\bibfnamefont {M.}~\bibnamefont {Coreno}}, \bibinfo {author} {\bibfnamefont {A.}~\bibnamefont {Demidovich}}, \bibinfo {author} {\bibfnamefont {M.}~\bibnamefont {Devetta}}, \bibinfo {author} {\bibfnamefont {M.}~\bibnamefont {Di~Fraia}}, \bibinfo {author} {\bibfnamefont {M.}~\bibnamefont {Drabbels}}, \emph {et~al.},\ }\bibfield  {title} {\bibinfo {title} {Ultrafast relaxation of photoexcited superfluid {He} nanodroplets},\ }\href@noop {} {\bibfield  {journal} {\bibinfo  {journal} {Nat. Commun.}\ }\textbf {\bibinfo {volume} {11}} (\bibinfo {year} {2020})}\BibitemShut {NoStop}%
\bibitem [{\citenamefont {Potts}\ and\ \citenamefont {Price}(1977)}]{potts1977photoelectron}%
  \BibitemOpen
  \bibfield  {author} {\bibinfo {author} {\bibfnamefont {A.}~\bibnamefont {Potts}}\ and\ \bibinfo {author} {\bibfnamefont {W.}~\bibnamefont {Price}},\ }\bibfield  {title} {\bibinfo {title} {Photoelectron studies of ionic materials using molecular beam techniques},\ }\href@noop {} {\bibfield  {journal} {\bibinfo  {journal} {Phys. Scr.}\ }\textbf {\bibinfo {volume} {16}},\ \bibinfo {pages} {191} (\bibinfo {year} {1977})}\BibitemShut {NoStop}%
\bibitem [{\citenamefont {Asmussen}\ \emph {et~al.}(2021)\citenamefont {Asmussen}, \citenamefont {Michiels}, \citenamefont {Dulitz}, \citenamefont {Ngai}, \citenamefont {Bangert}, \citenamefont {Barranco}, \citenamefont {Binz}, \citenamefont {Bruder}, \citenamefont {Danailov}, \citenamefont {Di~Fraia}, \citenamefont {Eloranta}, \citenamefont {Feifel}, \citenamefont {Giannessi}, \citenamefont {Pi}, \citenamefont {Plekan}, \citenamefont {Prince}, \citenamefont {Squibb}, \citenamefont {Uhl}, \citenamefont {Wituschek}, \citenamefont {Zangrando}, \citenamefont {Callegari}, \citenamefont {Stienkemeier},\ and\ \citenamefont {Mudrich}}]{Asmussen:2021}%
  \BibitemOpen
  \bibfield  {author} {\bibinfo {author} {\bibfnamefont {J.~D.}\ \bibnamefont {Asmussen}}, \bibinfo {author} {\bibfnamefont {R.}~\bibnamefont {Michiels}}, \bibinfo {author} {\bibfnamefont {K.}~\bibnamefont {Dulitz}}, \bibinfo {author} {\bibfnamefont {A.}~\bibnamefont {Ngai}}, \bibinfo {author} {\bibfnamefont {U.}~\bibnamefont {Bangert}}, \bibinfo {author} {\bibfnamefont {M.}~\bibnamefont {Barranco}}, \bibinfo {author} {\bibfnamefont {M.}~\bibnamefont {Binz}}, \bibinfo {author} {\bibfnamefont {L.}~\bibnamefont {Bruder}}, \bibinfo {author} {\bibfnamefont {M.}~\bibnamefont {Danailov}}, \bibinfo {author} {\bibfnamefont {M.}~\bibnamefont {Di~Fraia}}, \bibinfo {author} {\bibfnamefont {J.}~\bibnamefont {Eloranta}}, \bibinfo {author} {\bibfnamefont {R.}~\bibnamefont {Feifel}}, \bibinfo {author} {\bibfnamefont {L.}~\bibnamefont {Giannessi}}, \bibinfo {author} {\bibfnamefont {M.}~\bibnamefont {Pi}}, \bibinfo {author} {\bibfnamefont {O.}~\bibnamefont {Plekan}}, \bibinfo {author} {\bibfnamefont {K.~C.}\ \bibnamefont
  {Prince}}, \bibinfo {author} {\bibfnamefont {R.~J.}\ \bibnamefont {Squibb}}, \bibinfo {author} {\bibfnamefont {D.}~\bibnamefont {Uhl}}, \bibinfo {author} {\bibfnamefont {A.}~\bibnamefont {Wituschek}}, \bibinfo {author} {\bibfnamefont {M.}~\bibnamefont {Zangrando}}, \bibinfo {author} {\bibfnamefont {C.}~\bibnamefont {Callegari}}, \bibinfo {author} {\bibfnamefont {F.}~\bibnamefont {Stienkemeier}},\ and\ \bibinfo {author} {\bibfnamefont {M.}~\bibnamefont {Mudrich}},\ }\bibfield  {title} {\bibinfo {title} {Unravelling the full relaxation dynamics of superexcited helium nanodroplets},\ }\href@noop {} {\bibfield  {journal} {\bibinfo  {journal} {Phys. Chem. Chem. Phys.}\ }\textbf {\bibinfo {volume} {23}},\ \bibinfo {pages} {15138} (\bibinfo {year} {2021})}\BibitemShut {NoStop}%
\bibitem [{\citenamefont {Asmussen}\ \emph {et~al.}(2023{\natexlab{b}})\citenamefont {Asmussen}, \citenamefont {Ben~Ltaief}, \citenamefont {Sishodia}, \citenamefont {Abid}, \citenamefont {Bastian}, \citenamefont {Krishnan}, \citenamefont {Pedersen},\ and\ \citenamefont {Mudrich}}]{asmussen2023dopant}%
  \BibitemOpen
  \bibfield  {author} {\bibinfo {author} {\bibfnamefont {J.~D.}\ \bibnamefont {Asmussen}}, \bibinfo {author} {\bibfnamefont {L.}~\bibnamefont {Ben~Ltaief}}, \bibinfo {author} {\bibfnamefont {K.}~\bibnamefont {Sishodia}}, \bibinfo {author} {\bibfnamefont {A.~R.}\ \bibnamefont {Abid}}, \bibinfo {author} {\bibfnamefont {B.}~\bibnamefont {Bastian}}, \bibinfo {author} {\bibfnamefont {S.}~\bibnamefont {Krishnan}}, \bibinfo {author} {\bibfnamefont {H.~B.}\ \bibnamefont {Pedersen}},\ and\ \bibinfo {author} {\bibfnamefont {M.}~\bibnamefont {Mudrich}},\ }\bibfield  {title} {\bibinfo {title} {Dopant ionization and efficiency of ion and electron ejection from helium nanodroplets},\ }\href@noop {} {\bibfield  {journal} {\bibinfo  {journal} {J. Chem. Phys.}\ }\textbf {\bibinfo {volume} {159}} (\bibinfo {year} {2023}{\natexlab{b}})}\BibitemShut {NoStop}%
\bibitem [{\citenamefont {Sishodia}\ \emph {et~al.}(2025)\citenamefont {Sishodia}, \citenamefont {Ltaief}, \citenamefont {Scheel}, \citenamefont {Földes}, \citenamefont {Roos}, \citenamefont {Albrecht}, \citenamefont {Staněk}, \citenamefont {Jurkovičová}, \citenamefont {Hort}, \citenamefont {Nejdl}, \citenamefont {García-Alfonso}, \citenamefont {Halberstadt}, \citenamefont {Andreasson}, \citenamefont {Klimešová}, \citenamefont {Krikunova}, \citenamefont {Krishnan}, \citenamefont {Heidenreich},\ and\ \citenamefont {Mudrich}}]{sishodia2025xuvfluorescenceprobeinteratomic}%
  \BibitemOpen
  \bibfield  {author} {\bibinfo {author} {\bibfnamefont {K.}~\bibnamefont {Sishodia}}, \bibinfo {author} {\bibfnamefont {L.~B.}\ \bibnamefont {Ltaief}}, \bibinfo {author} {\bibfnamefont {N.}~\bibnamefont {Scheel}}, \bibinfo {author} {\bibfnamefont {I.~B.}\ \bibnamefont {Földes}}, \bibinfo {author} {\bibfnamefont {A.~H.}\ \bibnamefont {Roos}}, \bibinfo {author} {\bibfnamefont {M.}~\bibnamefont {Albrecht}}, \bibinfo {author} {\bibfnamefont {M.}~\bibnamefont {Staněk}}, \bibinfo {author} {\bibfnamefont {L.}~\bibnamefont {Jurkovičová}}, \bibinfo {author} {\bibfnamefont {O.}~\bibnamefont {Hort}}, \bibinfo {author} {\bibfnamefont {J.}~\bibnamefont {Nejdl}}, \bibinfo {author} {\bibfnamefont {E.}~\bibnamefont {García-Alfonso}}, \bibinfo {author} {\bibfnamefont {N.}~\bibnamefont {Halberstadt}}, \bibinfo {author} {\bibfnamefont {J.}~\bibnamefont {Andreasson}}, \bibinfo {author} {\bibfnamefont {E.}~\bibnamefont {Klimešová}}, \bibinfo {author} {\bibfnamefont {M.}~\bibnamefont {Krikunova}}, \bibinfo {author}
  {\bibfnamefont {S.}~\bibnamefont {Krishnan}}, \bibinfo {author} {\bibfnamefont {A.}~\bibnamefont {Heidenreich}},\ and\ \bibinfo {author} {\bibfnamefont {M.}~\bibnamefont {Mudrich}},\ }\href {https://arxiv.org/abs/2509.09532} {\bibinfo {title} {\mbox{XUV} fluorescence as a probe of interatomic \uppercase{C}oulombic decay of resonantly excited \uppercase{H}e nanodroplets}} (\bibinfo {year} {2025}),\ \Eprint {https://arxiv.org/abs/2509.09532} {arXiv:2509.09532 [physics.atm-clus]} \BibitemShut {NoStop}%
\bibitem [{\citenamefont {Asmussen}\ \emph {et~al.}(2023{\natexlab{c}})\citenamefont {Asmussen}, \citenamefont {Sishodia}, \citenamefont {Bastian}, \citenamefont {Abid}, \citenamefont {Ltaief}, \citenamefont {Pedersen}, \citenamefont {De}, \citenamefont {Medina}, \citenamefont {Pal}, \citenamefont {Richter} \emph {et~al.}}]{asmussen2023electron}%
  \BibitemOpen
  \bibfield  {author} {\bibinfo {author} {\bibfnamefont {J.~D.}\ \bibnamefont {Asmussen}}, \bibinfo {author} {\bibfnamefont {K.}~\bibnamefont {Sishodia}}, \bibinfo {author} {\bibfnamefont {B.}~\bibnamefont {Bastian}}, \bibinfo {author} {\bibfnamefont {A.~R.}\ \bibnamefont {Abid}}, \bibinfo {author} {\bibfnamefont {L.~B.}\ \bibnamefont {Ltaief}}, \bibinfo {author} {\bibfnamefont {H.~B.}\ \bibnamefont {Pedersen}}, \bibinfo {author} {\bibfnamefont {S.}~\bibnamefont {De}}, \bibinfo {author} {\bibfnamefont {C.}~\bibnamefont {Medina}}, \bibinfo {author} {\bibfnamefont {N.}~\bibnamefont {Pal}}, \bibinfo {author} {\bibfnamefont {R.}~\bibnamefont {Richter}}, \emph {et~al.},\ }\bibfield  {title} {\bibinfo {title} {Electron energy loss and angular asymmetry induced by elastic scattering in superfluid helium nanodroplets},\ }\href@noop {} {\bibfield  {journal} {\bibinfo  {journal} {Nanoscale}\ }\textbf {\bibinfo {volume} {15}},\ \bibinfo {pages} {14025} (\bibinfo {year} {2023}{\natexlab{c}})}\BibitemShut {NoStop}%
\bibitem [{\citenamefont {Ren}\ and\ \citenamefont {Kresin}(2007)}]{ren2007surface}%
  \BibitemOpen
  \bibfield  {author} {\bibinfo {author} {\bibfnamefont {Y.}~\bibnamefont {Ren}}\ and\ \bibinfo {author} {\bibfnamefont {V.~V.}\ \bibnamefont {Kresin}},\ }\bibfield  {title} {\bibinfo {title} {Surface location of alkaline-earth-metal-atom impurities on helium nanodroplets},\ }\href@noop {} {\bibfield  {journal} {\bibinfo  {journal} {Phys. Rev. A}\ }\textbf {\bibinfo {volume} {76}},\ \bibinfo {pages} {043204} (\bibinfo {year} {2007})}\BibitemShut {NoStop}%
\bibitem [{\citenamefont {Thaler}\ \emph {et~al.}(2020)\citenamefont {Thaler}, \citenamefont {Heim}, \citenamefont {Treiber},\ and\ \citenamefont {Koch}}]{thaler2020ultrafast}%
  \BibitemOpen
  \bibfield  {author} {\bibinfo {author} {\bibfnamefont {B.}~\bibnamefont {Thaler}}, \bibinfo {author} {\bibfnamefont {P.}~\bibnamefont {Heim}}, \bibinfo {author} {\bibfnamefont {L.}~\bibnamefont {Treiber}},\ and\ \bibinfo {author} {\bibfnamefont {M.}~\bibnamefont {Koch}},\ }\bibfield  {title} {\bibinfo {title} {Ultrafast photoinduced dynamics of single atoms solvated inside helium nanodroplets},\ }\href@noop {} {\bibfield  {journal} {\bibinfo  {journal} {J. Chem. Phys.}\ }\textbf {\bibinfo {volume} {152}} (\bibinfo {year} {2020})}\BibitemShut {NoStop}%
\bibitem [{\citenamefont {Kuma}\ \emph {et~al.}(2007)\citenamefont {Kuma}, \citenamefont {Goto}, \citenamefont {Slipchenko}, \citenamefont {Vilesov}, \citenamefont {Khramov},\ and\ \citenamefont {Momose}}]{kuma2007laser}%
  \BibitemOpen
  \bibfield  {author} {\bibinfo {author} {\bibfnamefont {S.}~\bibnamefont {Kuma}}, \bibinfo {author} {\bibfnamefont {H.}~\bibnamefont {Goto}}, \bibinfo {author} {\bibfnamefont {M.~N.}\ \bibnamefont {Slipchenko}}, \bibinfo {author} {\bibfnamefont {A.~F.}\ \bibnamefont {Vilesov}}, \bibinfo {author} {\bibfnamefont {A.}~\bibnamefont {Khramov}},\ and\ \bibinfo {author} {\bibfnamefont {T.}~\bibnamefont {Momose}},\ }\bibfield  {title} {\bibinfo {title} {Laser induced fluorescence of {Mg}-phthalocyanine in {He} droplets: Evidence for fluxionality of large {H}$_2$ clusters at 0.38 {K}},\ }\href@noop {} {\bibfield  {journal} {\bibinfo  {journal} {J. Chem. Phys.}\ }\textbf {\bibinfo {volume} {127}},\ \bibinfo {pages} {214301} (\bibinfo {year} {2007})}\BibitemShut {NoStop}%
\bibitem [{\citenamefont {De}\ \emph {et~al.}(2024)\citenamefont {De}, \citenamefont {Abid}, \citenamefont {Asmussen}, \citenamefont {Ben~Ltaief}, \citenamefont {Sishodia}, \citenamefont {Ulmer}, \citenamefont {Pedersen}, \citenamefont {Krishnan},\ and\ \citenamefont {Mudrich}}]{De:2024}%
  \BibitemOpen
  \bibfield  {author} {\bibinfo {author} {\bibfnamefont {S.}~\bibnamefont {De}}, \bibinfo {author} {\bibfnamefont {A.}~\bibnamefont {Abid}}, \bibinfo {author} {\bibfnamefont {J.}~\bibnamefont {Asmussen}}, \bibinfo {author} {\bibfnamefont {L.}~\bibnamefont {Ben~Ltaief}}, \bibinfo {author} {\bibfnamefont {K.}~\bibnamefont {Sishodia}}, \bibinfo {author} {\bibfnamefont {A.}~\bibnamefont {Ulmer}}, \bibinfo {author} {\bibfnamefont {H.}~\bibnamefont {Pedersen}}, \bibinfo {author} {\bibfnamefont {S.}~\bibnamefont {Krishnan}},\ and\ \bibinfo {author} {\bibfnamefont {M.}~\bibnamefont {Mudrich}},\ }\bibfield  {title} {\bibinfo {title} {Fragmentation of water clusters formed in helium nanodroplets by charge transfer and penning ionization},\ }\href@noop {} {\bibfield  {journal} {\bibinfo  {journal} {J. Chem. Phys.}\ }\textbf {\bibinfo {volume} {160}},\ \bibinfo {pages} {094308} (\bibinfo {year} {2024})}\BibitemShut {NoStop}%
\bibitem [{\citenamefont {Shi}\ \emph {et~al.}(2023)\citenamefont {Shi}, \citenamefont {Wang}, \citenamefont {Li},\ and\ \citenamefont {Su}}]{shi23}%
  \BibitemOpen
  \bibfield  {author} {\bibinfo {author} {\bibfnamefont {Y.}~\bibnamefont {Shi}}, \bibinfo {author} {\bibfnamefont {P.}~\bibnamefont {Wang}}, \bibinfo {author} {\bibfnamefont {W.}~\bibnamefont {Li}},\ and\ \bibinfo {author} {\bibfnamefont {Y.}~\bibnamefont {Su}},\ }\bibfield  {title} {\bibinfo {title} {Effects of \uppercase{N}a$^+$ and \uppercase{C}l$^-$ on hydrated clusters by {\em ab initio} study},\ }\href@noop {} {\bibfield  {journal} {\bibinfo  {journal} {J. Chem. Phys.}\ }\textbf {\bibinfo {volume} {159}},\ \bibinfo {pages} {044305} (\bibinfo {year} {2023})}\BibitemShut {NoStop}%
\bibitem [{\citenamefont {Sch{\"o}bel}\ \emph {et~al.}(2010)\citenamefont {Sch{\"o}bel}, \citenamefont {Bartl}, \citenamefont {Leidlmair}, \citenamefont {Daxner}, \citenamefont {Z{\"o}ttl}, \citenamefont {Denifl}, \citenamefont {M{\"a}rk}, \citenamefont {Scheier}, \citenamefont {Sp{\aa}ngberg}, \citenamefont {Mauracher} \emph {et~al.}}]{schobel2010sequential}%
  \BibitemOpen
  \bibfield  {author} {\bibinfo {author} {\bibfnamefont {H.}~\bibnamefont {Sch{\"o}bel}}, \bibinfo {author} {\bibfnamefont {P.}~\bibnamefont {Bartl}}, \bibinfo {author} {\bibfnamefont {C.}~\bibnamefont {Leidlmair}}, \bibinfo {author} {\bibfnamefont {M.}~\bibnamefont {Daxner}}, \bibinfo {author} {\bibfnamefont {S.}~\bibnamefont {Z{\"o}ttl}}, \bibinfo {author} {\bibfnamefont {S.}~\bibnamefont {Denifl}}, \bibinfo {author} {\bibfnamefont {T.~D.}\ \bibnamefont {M{\"a}rk}}, \bibinfo {author} {\bibfnamefont {P.}~\bibnamefont {Scheier}}, \bibinfo {author} {\bibfnamefont {D.}~\bibnamefont {Sp{\aa}ngberg}}, \bibinfo {author} {\bibfnamefont {A.}~\bibnamefont {Mauracher}}, \emph {et~al.},\ }\bibfield  {title} {\bibinfo {title} {Sequential penning ionization: harvesting energy with ions},\ }\href@noop {} {\bibfield  {journal} {\bibinfo  {journal} {Phys. Rev. Lett.}\ }\textbf {\bibinfo {volume} {105}},\ \bibinfo {pages} {243402} (\bibinfo {year} {2010})}\BibitemShut {NoStop}%
\bibitem [{\citenamefont {Foitzik}\ \emph {et~al.}(2025)\citenamefont {Foitzik}, \citenamefont {Bartolomei}, \citenamefont {Campos-Martínez}, \citenamefont {Pirani}, \citenamefont {Bergmeister}, \citenamefont {Ganner}, \citenamefont {Stromberg}, \citenamefont {Zappa}, \citenamefont {Mahmoodi-Darian}, \citenamefont {González-Lezana},\ and\ \citenamefont {Gruber}}]{Foitzik_Large_Ordered_Helium_Solvation_Shells}%
  \BibitemOpen
  \bibfield  {author} {\bibinfo {author} {\bibfnamefont {F.}~\bibnamefont {Foitzik}}, \bibinfo {author} {\bibfnamefont {M.}~\bibnamefont {Bartolomei}}, \bibinfo {author} {\bibfnamefont {J.}~\bibnamefont {Campos-Martínez}}, \bibinfo {author} {\bibfnamefont {F.}~\bibnamefont {Pirani}}, \bibinfo {author} {\bibfnamefont {S.}~\bibnamefont {Bergmeister}}, \bibinfo {author} {\bibfnamefont {L.}~\bibnamefont {Ganner}}, \bibinfo {author} {\bibfnamefont {I.}~\bibnamefont {Stromberg}}, \bibinfo {author} {\bibfnamefont {F.}~\bibnamefont {Zappa}}, \bibinfo {author} {\bibfnamefont {M.}~\bibnamefont {Mahmoodi-Darian}}, \bibinfo {author} {\bibfnamefont {T.}~\bibnamefont {González-Lezana}},\ and\ \bibinfo {author} {\bibfnamefont {E.}~\bibnamefont {Gruber}},\ }\bibfield  {title} {\bibinfo {title} {Formation of large ordered helium solvation shells around multiply charged species: \mbox{He$_N$Ho$^{2+}$} and \mbox{He$_N$Bi$^{2+}$}},\ }\href@noop {} {\bibfield  {journal} {\bibinfo  {journal} {Small Structures}\ }\textbf
  {\bibinfo {volume} {6}},\ \bibinfo {pages} {2500094} (\bibinfo {year} {2025})}\BibitemShut {NoStop}%
\bibitem [{\citenamefont {B{\"u}nermann}\ and\ \citenamefont {Stienkemeier}(2011)}]{Buenermann:2011}%
  \BibitemOpen
  \bibfield  {author} {\bibinfo {author} {\bibfnamefont {O.}~\bibnamefont {B{\"u}nermann}}\ and\ \bibinfo {author} {\bibfnamefont {F.}~\bibnamefont {Stienkemeier}},\ }\bibfield  {title} {\bibinfo {title} {Modeling the formation of alkali clusters attached to helium nanodroplets and the abundance of high-spin states},\ }\href@noop {} {\bibfield  {journal} {\bibinfo  {journal} {Eur. Phys. J. D}\ }\textbf {\bibinfo {volume} {61}},\ \bibinfo {pages} {645} (\bibinfo {year} {2011})}\BibitemShut {NoStop}%
\bibitem [{\citenamefont {Joppien}\ \emph {et~al.}(1993)\citenamefont {Joppien}, \citenamefont {Karnbach},\ and\ \citenamefont {M\"oller}}]{Joppien:1993}%
  \BibitemOpen
  \bibfield  {author} {\bibinfo {author} {\bibfnamefont {M.}~\bibnamefont {Joppien}}, \bibinfo {author} {\bibfnamefont {R.}~\bibnamefont {Karnbach}},\ and\ \bibinfo {author} {\bibfnamefont {T.}~\bibnamefont {M\"oller}},\ }\bibfield  {title} {\bibinfo {title} {Electronic excitations in liquid helium: The evolution from small clusters to large droplets},\ }\href@noop {} {\bibfield  {journal} {\bibinfo  {journal} {Phys. Rev. Lett.}\ }\textbf {\bibinfo {volume} {71}},\ \bibinfo {pages} {2654} (\bibinfo {year} {1993})}\BibitemShut {NoStop}%
\bibitem [{\citenamefont {Nauta}\ and\ \citenamefont {Miller}(1999)}]{nauta1999}%
  \BibitemOpen
  \bibfield  {author} {\bibinfo {author} {\bibfnamefont {K.}~\bibnamefont {Nauta}}\ and\ \bibinfo {author} {\bibfnamefont {R.}~\bibnamefont {Miller}},\ }\bibfield  {title} {\bibinfo {title} {Nonequilibrium self-assembly of long-chains of polar molecules in superfluid helium},\ }\href@noop {} {\bibfield  {journal} {\bibinfo  {journal} {Science}\ }\textbf {\bibinfo {volume} {283}},\ \bibinfo {pages} {1895} (\bibinfo {year} {1999})}\BibitemShut {NoStop}%
\bibitem [{\citenamefont {Wang}\ \emph {et~al.}(2000)\citenamefont {Wang}, \citenamefont {Cieplak},\ and\ \citenamefont {Kollman}}]{wang00}%
  \BibitemOpen
  \bibfield  {author} {\bibinfo {author} {\bibfnamefont {J.}~\bibnamefont {Wang}}, \bibinfo {author} {\bibfnamefont {P.}~\bibnamefont {Cieplak}},\ and\ \bibinfo {author} {\bibfnamefont {K.~A.}\ \bibnamefont {Kollman}},\ }\bibfield  {title} {\bibinfo {title} {How does a restrained electrostatic potential (\uppercase{RESP}) model perform in calculating conformational energies of organic and biological molecules},\ }\href@noop {} {\bibfield  {journal} {\bibinfo  {journal} {J. Comput. Chem.}\ }\textbf {\bibinfo {volume} {21}},\ \bibinfo {pages} {1049} (\bibinfo {year} {2000})}\BibitemShut {NoStop}%
\bibitem [{\citenamefont {Frisch}\ \emph {et~al.}(2016)\citenamefont {Frisch}, \citenamefont {Trucks}, \citenamefont {Schlegel}, \citenamefont {Scuseria}, \citenamefont {Robb}, \citenamefont {Cheeseman}, \citenamefont {Scalmani}, \citenamefont {Barone}, \citenamefont {Petersson}, \citenamefont {Nakatsuji}, \citenamefont {Li}, \citenamefont {Caricato}, \citenamefont {Marenich}, \citenamefont {Bloino}, \citenamefont {Janesko}, \citenamefont {Gomperts}, \citenamefont {Mennucci}, \citenamefont {Hratchian}, \citenamefont {Ortiz}, \citenamefont {Izmaylov}, \citenamefont {Sonnenberg}, \citenamefont {Williams-Young}, \citenamefont {Ding}, \citenamefont {Lipparini}, \citenamefont {Egidi}, \citenamefont {Goings}, \citenamefont {Peng}, \citenamefont {Petrone}, \citenamefont {Henderson}, \citenamefont {Ranasinghe}, \citenamefont {Zakrzewski}, \citenamefont {Gao}, \citenamefont {Rega}, \citenamefont {Zheng}, \citenamefont {Liang}, \citenamefont {Hada}, \citenamefont {Ehara}, \citenamefont {Toyota}, \citenamefont {Fukuda},
  \citenamefont {Hasegawa}, \citenamefont {Ishida}, \citenamefont {Nakajima}, \citenamefont {Honda}, \citenamefont {Kitao}, \citenamefont {Nakai}, \citenamefont {Vreven}, \citenamefont {Throssell}, \citenamefont {Montgomery}, \citenamefont {Peralta}, \citenamefont {Ogliaro}, \citenamefont {Bearpark}, \citenamefont {Heyd}, \citenamefont {Brothers}, \citenamefont {Kudin}, \citenamefont {Staroverov}, \citenamefont {Keith}, \citenamefont {Kobayashi}, \citenamefont {Normand}, \citenamefont {Raghavachari}, \citenamefont {Rendell}, \citenamefont {Burant}, \citenamefont {Iyengar}, \citenamefont {Tomasi}, \citenamefont {Cossi}, \citenamefont {Millam}, \citenamefont {Klene}, \citenamefont {Adamo}, \citenamefont {Cammi}, \citenamefont {Ochterski}, \citenamefont {Martin}, \citenamefont {Morokuma}, \citenamefont {Farkas}, \citenamefont {Foresman},\ and\ \citenamefont {Fox}}]{g16}%
  \BibitemOpen
  \bibfield  {author} {\bibinfo {author} {\bibfnamefont {M.~J.}\ \bibnamefont {Frisch}}, \bibinfo {author} {\bibfnamefont {G.~W.}\ \bibnamefont {Trucks}}, \bibinfo {author} {\bibfnamefont {H.~B.}\ \bibnamefont {Schlegel}}, \bibinfo {author} {\bibfnamefont {G.~E.}\ \bibnamefont {Scuseria}}, \bibinfo {author} {\bibfnamefont {M.~A.}\ \bibnamefont {Robb}}, \bibinfo {author} {\bibfnamefont {J.~R.}\ \bibnamefont {Cheeseman}}, \bibinfo {author} {\bibfnamefont {G.}~\bibnamefont {Scalmani}}, \bibinfo {author} {\bibfnamefont {V.}~\bibnamefont {Barone}}, \bibinfo {author} {\bibfnamefont {G.~A.}\ \bibnamefont {Petersson}}, \bibinfo {author} {\bibfnamefont {H.}~\bibnamefont {Nakatsuji}}, \bibinfo {author} {\bibfnamefont {X.}~\bibnamefont {Li}}, \bibinfo {author} {\bibfnamefont {M.}~\bibnamefont {Caricato}}, \bibinfo {author} {\bibfnamefont {A.~V.}\ \bibnamefont {Marenich}}, \bibinfo {author} {\bibfnamefont {J.}~\bibnamefont {Bloino}}, \bibinfo {author} {\bibfnamefont {B.~G.}\ \bibnamefont {Janesko}}, \bibinfo {author}
  {\bibfnamefont {R.}~\bibnamefont {Gomperts}}, \bibinfo {author} {\bibfnamefont {B.}~\bibnamefont {Mennucci}}, \bibinfo {author} {\bibfnamefont {H.~P.}\ \bibnamefont {Hratchian}}, \bibinfo {author} {\bibfnamefont {J.~V.}\ \bibnamefont {Ortiz}}, \bibinfo {author} {\bibfnamefont {A.~F.}\ \bibnamefont {Izmaylov}}, \bibinfo {author} {\bibfnamefont {J.~L.}\ \bibnamefont {Sonnenberg}}, \bibinfo {author} {\bibfnamefont {D.}~\bibnamefont {Williams-Young}}, \bibinfo {author} {\bibfnamefont {F.}~\bibnamefont {Ding}}, \bibinfo {author} {\bibfnamefont {F.}~\bibnamefont {Lipparini}}, \bibinfo {author} {\bibfnamefont {F.}~\bibnamefont {Egidi}}, \bibinfo {author} {\bibfnamefont {J.}~\bibnamefont {Goings}}, \bibinfo {author} {\bibfnamefont {B.}~\bibnamefont {Peng}}, \bibinfo {author} {\bibfnamefont {A.}~\bibnamefont {Petrone}}, \bibinfo {author} {\bibfnamefont {T.}~\bibnamefont {Henderson}}, \bibinfo {author} {\bibfnamefont {D.}~\bibnamefont {Ranasinghe}}, \bibinfo {author} {\bibfnamefont {V.~G.}\ \bibnamefont
  {Zakrzewski}}, \bibinfo {author} {\bibfnamefont {J.}~\bibnamefont {Gao}}, \bibinfo {author} {\bibfnamefont {N.}~\bibnamefont {Rega}}, \bibinfo {author} {\bibfnamefont {G.}~\bibnamefont {Zheng}}, \bibinfo {author} {\bibfnamefont {W.}~\bibnamefont {Liang}}, \bibinfo {author} {\bibfnamefont {M.}~\bibnamefont {Hada}}, \bibinfo {author} {\bibfnamefont {M.}~\bibnamefont {Ehara}}, \bibinfo {author} {\bibfnamefont {K.}~\bibnamefont {Toyota}}, \bibinfo {author} {\bibfnamefont {R.}~\bibnamefont {Fukuda}}, \bibinfo {author} {\bibfnamefont {J.}~\bibnamefont {Hasegawa}}, \bibinfo {author} {\bibfnamefont {M.}~\bibnamefont {Ishida}}, \bibinfo {author} {\bibfnamefont {T.}~\bibnamefont {Nakajima}}, \bibinfo {author} {\bibfnamefont {Y.}~\bibnamefont {Honda}}, \bibinfo {author} {\bibfnamefont {O.}~\bibnamefont {Kitao}}, \bibinfo {author} {\bibfnamefont {H.}~\bibnamefont {Nakai}}, \bibinfo {author} {\bibfnamefont {T.}~\bibnamefont {Vreven}}, \bibinfo {author} {\bibfnamefont {K.}~\bibnamefont {Throssell}}, \bibinfo {author}
  {\bibfnamefont {J.~A.}\ \bibnamefont {Montgomery}, \bibfnamefont {{Jr.}}}, \bibinfo {author} {\bibfnamefont {J.~E.}\ \bibnamefont {Peralta}}, \bibinfo {author} {\bibfnamefont {F.}~\bibnamefont {Ogliaro}}, \bibinfo {author} {\bibfnamefont {M.~J.}\ \bibnamefont {Bearpark}}, \bibinfo {author} {\bibfnamefont {J.~J.}\ \bibnamefont {Heyd}}, \bibinfo {author} {\bibfnamefont {E.~N.}\ \bibnamefont {Brothers}}, \bibinfo {author} {\bibfnamefont {K.~N.}\ \bibnamefont {Kudin}}, \bibinfo {author} {\bibfnamefont {V.~N.}\ \bibnamefont {Staroverov}}, \bibinfo {author} {\bibfnamefont {T.~A.}\ \bibnamefont {Keith}}, \bibinfo {author} {\bibfnamefont {R.}~\bibnamefont {Kobayashi}}, \bibinfo {author} {\bibfnamefont {J.}~\bibnamefont {Normand}}, \bibinfo {author} {\bibfnamefont {K.}~\bibnamefont {Raghavachari}}, \bibinfo {author} {\bibfnamefont {A.~P.}\ \bibnamefont {Rendell}}, \bibinfo {author} {\bibfnamefont {J.~C.}\ \bibnamefont {Burant}}, \bibinfo {author} {\bibfnamefont {S.~S.}\ \bibnamefont {Iyengar}}, \bibinfo {author}
  {\bibfnamefont {J.}~\bibnamefont {Tomasi}}, \bibinfo {author} {\bibfnamefont {M.}~\bibnamefont {Cossi}}, \bibinfo {author} {\bibfnamefont {J.~M.}\ \bibnamefont {Millam}}, \bibinfo {author} {\bibfnamefont {M.}~\bibnamefont {Klene}}, \bibinfo {author} {\bibfnamefont {C.}~\bibnamefont {Adamo}}, \bibinfo {author} {\bibfnamefont {R.}~\bibnamefont {Cammi}}, \bibinfo {author} {\bibfnamefont {J.~W.}\ \bibnamefont {Ochterski}}, \bibinfo {author} {\bibfnamefont {R.~L.}\ \bibnamefont {Martin}}, \bibinfo {author} {\bibfnamefont {K.}~\bibnamefont {Morokuma}}, \bibinfo {author} {\bibfnamefont {O.}~\bibnamefont {Farkas}}, \bibinfo {author} {\bibfnamefont {J.~B.}\ \bibnamefont {Foresman}},\ and\ \bibinfo {author} {\bibfnamefont {D.~J.}\ \bibnamefont {Fox}},\ }\href@noop {} {\bibinfo {title} {Gaussian˜16 {R}evision {C}.01}} (\bibinfo {year} {2016}),\ \bibinfo {note} {gaussian Inc. Wallingford CT}\BibitemShut {NoStop}%
\bibitem [{\citenamefont {Bacchus-Montabonel}\ and\ \citenamefont {Calvo}(2015)}]{mcbmuracil}%
  \BibitemOpen
  \bibfield  {author} {\bibinfo {author} {\bibfnamefont {M.-C.}\ \bibnamefont {Bacchus-Montabonel}}\ and\ \bibinfo {author} {\bibfnamefont {F.}~\bibnamefont {Calvo}},\ }\bibfield  {title} {\bibinfo {title} {Nanohydration of uracil: emergence of three-dimensional structures and proton-induced charge transfer},\ }\href@noop {} {\bibfield  {journal} {\bibinfo  {journal} {Phys. Chem. Chem. Phys.}\ }\textbf {\bibinfo {volume} {17}},\ \bibinfo {pages} {9629} (\bibinfo {year} {2015})}\BibitemShut {NoStop}%
\end{thebibliography}%

\end{document}